**Interfacial strong coupling and negative dispersion of propagating polaritons in freestanding oxide membranes**

**Authors:** Brayden Lukaskawcez[1], Shivasheesh Varshney[2], Sooho Choo[2], Sang Hyun Park[3], Dongjea Seo[3], Liam Thompson[1], Nitzan Hirshberg[1], Madison Garber[1], Devon Uram[1], Hayden Binger[1], Steven Koester[3], Sang-Hyun Oh[3], Tony Low[3], Bharat Jalan[2], Alexander McLeod[1,*]

1. *School of Physics and Astronomy, University of Minnesota, Twin Cities, Minneapolis, Minnesota 55455, USA*
2. *Department of Chemical Engineering and Materials Science, University of Minnesota, Twin Cities, Minneapolis, Minnesota 55455, USA*
3. *Department of Electrical and Computer Engineering, University of Minnesota, Minneapolis, Minnesota 55455, USA*

* *Corresponding Author: mcleoda@umn.edu*

## Abstract:

Membranes of complex oxides like perovskite $SrTiO_3$ extend the multi-functional promise of oxide electronics into the nanoscale regime of two-dimensional materials. Here we demonstrate that free-standing oxide membranes supply a reconfigurable platform for nano-photonics based on propagating surface phonon polaritons. We apply infrared near-field imaging and -spectroscopy enabled by a tunable ultrafast laser to study pristine nano-thick $SrTiO_3$ membranes prepared by hybrid molecular beam epitaxy. As predicted by coupled mode theory, we find that strong coupling of interfacial polaritons realizes symmetric and antisymmetric hybridized modes with simultaneously tunable negative and positive group velocities. By resolving reflection of these propagating modes from membrane edges, defects, and substrate structures, we quantify their dispersion with position-resolved nano-spectroscopy. Remarkably, we find polariton negative dispersion is both robust and tunable through choice of membrane dielectric environment and thickness and propose a novel design for in-plane Veselago lensing harnessing this control. Our work lays the foundation for tunable transformation optics at the nanoscale using polaritons in a wide range of freestanding complex oxide membranes.

**[Main Text]**

Two-dimensional materials promise to enable wide-ranging applications for sensing, information processing, and light manipulation at deeply subwavelength scales. Polaritons, hybrid excitations of light and matter in these media, enable confining and manipulating electromagnetic energy at length-scales hundreds of times below the diffraction-limit:[1,2] $\lambda_p/\lambda \ll 100^{-1}$, where $\lambda_p = 2\pi/q_p$ is the polariton wavelength (and $q_p$ its momentum), and $\lambda = 2\pi c/\omega$ is the photon wavelength determined by the speed of light $c$ and frequency $\omega$. Recent studies[3–6] have uncovered polaritons whose energy dispersion $\omega_p(q)$ showcases negative group velocity $v_g = {d\omega_p}/{dq} < 0$, a phenomenon qualified as "negative (group) index of refraction" $n_g \equiv \frac{v_g}{c} < 0$ with antiparallel phase and energy transfer velocities. Juxtaposing media with positive and negative $n_g$ yields phenomena studied more than half a century ago by V. Veselago, including profound emergence of a "perfect (Veselago) lens".[7,8] More generally, unconventional subwavelength control of electromagnetic waves based on "transformation optics" call for media with negative index realized inhomogeneously on-demand.[9] Complex metamaterials realizing simultaneously negative effective permittivity and permeability have produced negative refraction, successfully elevating these proposals beyond theory,[10–12] though at limited photon energies, and restricted to supra-wavelength scales. Meanwhile, pursuits to engineer negative refraction for nano-confined light via polaritons remain in their infancy. Surface phonon polaritons (SPhPs) with "negative dispersion" are observed in type-I (including "in-plane") hyperbolic materials,[2,6,13–15] though the limited palette and tunability (e.g. towards $n_g > 0$) of eligible ionic crystals hampers functionalizing polariton negative dispersion, with notable exceptions.[3–5] We explore an alternative paradigm that exploits "conventional" polaritons in the unconventional medium of nano-thick membranes for expansive realizations of negative-index nano-photonics.

Here, combining material synthesis by hybrid molecular beam epitaxy (hMBE)[16] with characterization by infrared (IR) nano-spectroscopy (Fig. 1a-b), we demonstrate rational control of negative polariton dispersion at deeply subwavelength scales in an expansive new family of quasi-2D media: high-quality free-standing oxide membranes (Fig. 1e).[17–19] We demonstrate strong coupling of surface optical phonon polaritons across interfaces of nano-thick $SrTiO_3$ (STO)[20,21] realizes propagating membrane surface phonon polaritons (mSPhPs) with tunable negative dispersion. An archetypical complex oxide, STO exhibits a vibrant phase diagram including tunable metallic,[22] ferroelectric,[20,23] superconducting,[24,25] and multiferroic phases,[24] promising multifunctionality unavailable in conventional polaritonic media. While polariton negative refraction has been realized in wavelength-scale and heterostructured subwavelength media,[26,27] our study relies on strongly coupling polaritons across high-quality interfaces of a single membrane. Interfacial plasmon coupling is well-studied in the context of metallic thin films (*e.g.* insulator-metal-insulator waveguides),[28] but polariton coupling in thickness-controlled free-standing dielectrics, while acknowledged,[1,21] remains unexploited, and real-space propagation of polaritons with negative dispersion remains unresolved in this paradigm, till now. Our study tests the following elementary model of tunable negative dispersion (Fig. 1c): The interface separating a "metallic" membrane (permittivity $\varepsilon < 0$) from a dielectric substrate ($\varepsilon_{\rm sub}$) supports a surface optical phonon with momentum $q_p \gg \omega/c$ wherever $\varepsilon(\omega \equiv \omega_{SO}^b) = -\varepsilon_{\rm sub}$,[1,2,29] and so too does the membrane top surface (above which *e.g.* $\varepsilon = 1$) at $\omega \equiv \omega_{SO}^t$. Polaritons reside within the Restrahlen band ($\omega_{TO} < \omega_{SO} < \omega_{LO}$) of an infrared-active excitation; our work considers that associated with the highest energy STO optical phonon.[21,30,31]

Interfacial polaritons comprise a dynamical system with detuning $\delta\omega \equiv |\omega_{SO}^{t} - \omega_{SO}^{b}|$ that, owing to evanescent decay of polariton fields across the membrane thickness $d$, couple with a momentum-dependent Rabi frequency $\Omega(q) \equiv \Delta_0 e^{-qd}$, with $\Delta_0 \sim \sqrt{\omega_{LO}^2 - \omega_{TO}^2}$ a polariton oscillator strength. As illustrated in Fig. 1d, normal modes of the coupled dynamical system resonate as "symmetric" (at $\omega = \omega_+$) and "antisymmetric" ($\omega_-$) linear combinations[1,28,32,33] of the original polaritons with dispersions estimable by the "Rabi formula": $\omega_{\pm}(q) \approx \langle\omega_{SO}\rangle \mp \frac{1}{2}\sqrt{\delta\omega^2 + \Omega(q)^2}$ (with $\langle\omega_{SO}\rangle \equiv (\omega_{SO}^{t} + \omega_{SO}^{b})/2$ ).[34,35] These normal modes showcase conventional ($n_g^+ > 0$) and unconventional ($n_g^- < 0$) dispersions, respectively, in accord with detailed numerical predictions of the realistic loss function for mSPhPs in an STO membrane (Fig. 1b; details in Supplementary Information, *SI*).

Here, we apply energy-resolved near-field infrared nano-imaging and -spectroscopy to directly resolve $\omega_{\pm}(q)$ in high-quality STO membranes by polariton interferometry (Fig. 1b), confirming the hypothesis that membrane thickness and environment tune mSPhP dispersions through respective control of $\Omega$ and $\delta\omega$. Following these results, we 1) quantify the impact of membrane defects for mSPhP propagation, 2) establish mSPhPs as a platform for polariton strong coupling,[34,36] and 3) present a refined model to explain "anomalous" coupling exceeding predictions by the Rabi formula. 4) Finally, we propose a general scheme exploiting mSPhPs for Veselago nano-lensing[8] (illustrated in Fig. 1b) at application-scale in freestanding oxide membranes.

## Results

We prepare free-standing STO membranes of targeted thicknesses spanning d=15 $-$ 120nm through rapid hMBE growth[16] upon a water-soluble sacrificial layer, enabling large-area transfer at application-relevant scales (> 10mm) onto substrates chosen for control of polaritons (Fig. 1e; Methods). We employ scattering-type near-field optical microscopy and spectroscopy (nano-imaging and -spectroscopy) enabled by a single broadband tunable ultrafast laser source[37] and monochromator (Methods) to examine mSPhPs near edges and defects of these membranes spanning the mid-infrared ($\omega$ from 600-1000 cm$^{-1}$) with a typical spatial resolution of 10 nm (Methods). We begin our study with nano-spectroscopy across the Restrahlen band of a $d$ =25nm STO membrane transferred to $SiO_2$ substrate (Fig. 1c).[21,30,31,38] Fig. 2a presents the spectral reflectance (red) and absorption (blue) – the amplitude and phase of probe-scattered radiation generated by near-field interaction with the sample (Methods) – associated with the mSPhP response, consistent with our former results and affirming single-crystallinity.[16,30] Fig. 1b presents a triangular flake of this membrane imaged at discrete (monochromated) laser energies 660-750 cm$^{-1}$, revealing nano-scale absorption enhancement at membrane edges. Fig. 2d reveals the same phenomenon imaged at enhanced resolution at four uniformly incrementing energies in a similar ($d$=50nm) membrane flake which, for the first time with phase-resolved imaging, clearly resolve a negatively dispersing polariton in a complex oxide: The bright absorption fringe detected (e.g. at 750 cm$^{-1}$) along the flake edge results from "polariton interferometry" (Methods), whereby probe-launched polaritons reflect from a material edge and return to the probe, enhancing probe-scattered light at polariton round-trip travel distances yileding constructive interference (*cf.* Fig. 1b).[39–41] Line cuts of absorption across the flake (red lines) highlight mSPhP wavelengths increasing with photon energy, spanning the entire flake at 780 cm$^{-1}$, consistent with negative dispersion ($\frac{dq_p}{d\omega} < 0$). Fig. 2c compares $q_p$ obtained from each of these images (symbols; extraction described in Methods and ex in SI) with the loss function numerically predicted for this membrane (SI). Comparison with Fig. 1c affirms our detection of the negatively dispersing asymmetric mSPhP.

We now test predictions by the Rabi formula that negative dispersion of the asymmetric mode can be controlled through 1) substrate permittivity to detune ($\delta\omega$) interfacial polaritons, and 2) membrane thickness to regulate their coupling (Rabi frequency) $\Omega(q)$. We transfer our membranes of varied thicknesses to substrates of Si ($\varepsilon_{\text{sub}} = 11.7$), $SiO_2$ ($= 2.9$), and air ($= 1$) (combinations in Table S1); the latter by suspending membranes over etched Si (Methods). As sketched in Fig. 3a, we measure mSPhP dispersions by position-resolved ("line-scan") nano-spectroscopy (Methods).[41,42] We identify polariton interference from nano-resolved spatio-spectral amplitude and phase measurements by locating positions of maxima in polariton absorption relative to the membrane edge as a function of frequency (Methods). Figs. 3b,c present nano-spectroscopy across the edge of a suspended 50nm membrane, showcasing 10nm-resolved amplitude and phase of the mSPhP field as a function of position and energy over nearly the entire Restrahlen band. We observe both positive and negative ($\frac{d\lambda_p}{d\omega} <$ or $> 0$) dispersions at energies below and above $\omega = 740\ \text{cm}^{-1}$, respectively, as predicted by the loss function (dashed lines). We associate conventional and unconventional dispersions with symmetric and antisymmetric mSPhPs, respectively. Notably, dispersions "touch" (Fig. 3f) at $\omega_{SO} = 740\ \text{cm}^{-1}$ only where $\varepsilon_{\text{STO}}(\omega_{SO}) \approx -1$ and $\lambda_p \ll d$, signifying where interfacial polaritons decouple into degenerate modes ($\delta\omega = 0$) at top and bottom membrane interfaces, in accord with the simple coupled mode theory.

By applying this analysis to membranes of controlled thickness transferred to a common $SiO_2$ substrate we can now showcase that mSPhPs in our membranes exhibit equivalent tunability to that observed for phonon-polaritons in hBN[41]. Fig. 3e compares dispersions measured for the antisymmetric mSPhP with respective loss functions, showing agreement for 120nm and 55nm membranes, whereas for thinner membranes deviations emerge at photon energies approaching $\omega_{LO}$, arising perhaps from differences in optical phonon response from ultra-thin films compared with bulk STO.[30] Nevertheless, our study confirms that reducing membrane thickness "flattens" these dispersions, enabling control over the asymmetric mSPhP negative group velocity. For instance, we confirm that two-fold decrease in polariton wavelength and absolute group velocity $|v_g|$ both accompany reduction in membrane thickness by the same factor (e.g. 50nm to 25nm). This control owes to thickness-dependent interfacial coupling $\Omega(qd)$, whose increase (at fixed $q$) with decreasing $d$ predicts enhanced "mode repulsion" between symmetric and asymmetric mSPhPs, in accord with our observations. Next, we investigate detuning interfacial polaritons ($\delta\omega > 0$) through control of substrate permittivity $\varepsilon_{\text{sub}}$. Fig. 3f presents dispersions for a single 55nm membrane transferred onto contrasting substrates: uniform $SiO_2$ ($\varepsilon_{sub} = 3$) and etched Si with ~100nm deep trenches for suspending the membrane in vacuum ($\varepsilon_{sub}$=1). The supported membrane exhibits *longer* wavelength polaritons than the suspended one at all frequencies. Measurements of the same membrane transferred onto Si showcase further, albeit modest, increase in $\lambda_p$ at all energies (SI). Whereas studies of hBN, $MoO_3$, and graphene broadly find polariton wavelengths shorten in high-permittivity environments [43–45] consequent to decreasing $\omega_{SO}$, wavelengths of our antisymmetric mSPhP *grow* due to *decreased* mode repulsion of $\omega_-$ (in proportion to $\delta\omega$). Clearly, interfacial polariton coupling sustains unconventional negative dispersion tunable over a wide range of facile membrane configurations.

We now turn our attention to real-space tuning of the symmetric mSPhP at energies $\omega <$ 740 cm$^{-1}$. Fig. 3d (right) reveals a single 55nm $SrTiO_3$ flake (width $W = 600$nm) partially suspended over a 90nm deep trench (dashed white lines) in a silicon substrate. Imaging this membrane at 600 cm$^{-1}$ resolves apparent resonant absorption in the suspended (upper) region whereas mSPhPs appear absent from the supported (lower) region. To understand this behavior, we compare loss functions (left) predicted for the same membrane suspended (top) or supported by a silicon substrate (bottom). Clearly, the supported membrane shows an excitation gap $\Delta = \sqrt{\Omega^2 + \delta\omega^2} > \delta\omega \approx 150$ cm$^{-1}$ between symmetric and asymmetric polaritons that is minimized in the suspended membrane ($\Delta \sim \Omega(\mathrm{q})$). Thus at $\omega = 600$cm$^{-1}$ we expect symmetric mSPhPs "trapped" in the suspended membrane. Furthermore, here the loss function predicts polariton wavelength matching $\lambda_p \approx 2W$, explaining the confined "drumhead"-like polariton waveform apparent in both reflectivity and absorption maps. An additional fringe not predicted by the loss function is also visible along the entire membrane edge with equal periodicity in both the suspended and supported membrane. This fringe may indicate the presence of "edge mode polaritons" though further measurements are needed to confirm this.

Polaritons can also be impaired (or controlled) by membrane defects, as we now explore. Remarkably, transferring hMBE membranes as thin as 5nm leaves them largely intact over hundreds of microns (Fig. S7), though at expense of inhomogeneities like mechanical stresses and wrinkles (Fig. 4a). Such defects may back-scatter mSPhPs or diffract them into the far-field, depending on characteristic defect width $s$. We expect small defects ($s_{eff} << \lambda_p$) to transmit mSPhPs across their barrier, with partial reflection. To functionalize defects for device applications like mSPhP "beam-splitters", waveguides, and Fabry-Perot cavities, precise control over polariton propagation and scattering is crucial, motivating our study of $s$-dependent mSPhP scattering. Fig. 4a presents defects in our 55nm membrane identified by AFM topography, including 300nm (left) and 200nm-wide (middle) cracks, a 5nm tall wrinkle (right), and point defects approximately 1nm in height visible throughout. We perform nano-imaging of the antisymmetric mode ($\omega_-$ =750 cm$^{-1}$) near these defects (Fig. 4b-c). Along the 300nm wide crack (horizontal dark line), along the 5nm tall wrinkle (right), and surrounding each point-defect, we observe bright fringes associated with polariton reflection. For the wrinkle, the gradual separation of the membrane from the substrate occurs over approximately 250nm, and thus for the purposes of assigning a physical size relevant for scattering the polariton we make the assumption $s_{phys} = 250nm$. Fig. 4c highlights the ring-like shape of defect-scattered polaritons, analogous to plasmon scattering observed around charge defects in graphene monolayers.[46] We quantify mSPhP reflection from each of these defects across the dashed lines in Fig. 4a by line-scan nano-spectroscopy (Fig. 4d). In each measurement, bright absorption fringes on either side of each defect disperse with the same negative dispersion (dashes are guides to the eye) as observed at membrane edges. Constant-frequency cuts of these data measured across the 300nm crack (Fig. 4e) highlight the increasing wavelength (negative dispersion) of polaritons reflected from either side of the crack with increasing energy. Fig. 4d (red arrows) also highlights single point defects whose polariton reflections disperse like those from cracks, albeit with amplitude reduced to our detection limit. Color scales for Fig. 4d panels were adjusted to individually highlight polariton reflection from each defect. More quantitatively, Fig. 4f compares the measured amplitude of polariton reflection $R(s)$ at 7 values of polariton wavelength $\lambda_p(\omega)$ increasing across the bandwidth of negative mSPhP dispersion from four defects in Fig. 4d (identified by symbols), relative to reflection $R_{\mathrm{edge}}$ from the membrane edge (we take the latter as unity reflectance corresponding to infinite size $s$). In regards to scattering events of the polariton the only relevant length scale is $\lambda_p(\omega)$, thus employing a 1-D scattering model we find the squared polariton reflectance is well described by

$$\left(\frac{R}{R_{edge}}\right)^2 \sim 1 - A \exp\left(-\frac{2\,s}{\lambda_p(\omega)}\right)$$

(dashed curves) with $A \approx 1$, consistent with a tunneling process where the squared polariton transmission $T^2$ (unmeasured; proportional to $e^{-2s/\lambda_p}$) underlies reflectance $R^2 = 1 - T^2$. Best-fit values of $s_{eff}$ (Fig. 4f inset) accord with physical sizes of cracks and the wrinkle (Fig. 4a) to within a factor of 50%, whereas the point defect is exceptionally inefficient at scattering polaritons ($s_{eff} \ll 45$nm).

Finally, we explore mSPhP control via *engineered* confinement. Fig 4g presents a 25nm membrane transferred onto Si etched with ~100nm deep circular holes with $D$ =1 micron diameter. Here the membrane is purposefully suspended to confine the symmetric mSPhP at the resonance condition $\lambda_p \approx 2D$ at $\omega_+$ = 600 $cm^{-1}$ (the polariton is "gapped out" beyond the suspended diameter). Images reveal circular resonant absorption (right) within the suspended disk and no polariton absorption without. This "drumhead" for mSPhPs exhibits a half-wave peak in absorption (Fig. 4g, right), complemented by a ring in overall field amplitude (left) just inside the hole diameter, comparable to plasmon confinement achieved in graphene.[47] Further linescan spectroscopy of this disk-confined polariton is presented in SI.

## Discussion

Having demonstrated rational control over mSPhP dispersions, we now refine their quantitative description in comparison with the Rabi formula. We apply nano-spectroscopy of $d = 25$nm, 55nm, and 120nm STO membranes across their entire Restrahlen band to quantify dependence of symmetric-asymmetric mode splitting $\Delta(q_p) \equiv \omega_- - \omega_+$ on $\varepsilon_{sub}$ and $d$. Fig. 5a compares dispersions in suspended 25nm and 55nm membranes matching their respective loss functions. Here $\delta\omega = 0$, such that observing $\Delta = \Omega(q_p d)$ for fixed $q_p$ directly measures the $d$-dependence of interfacial polariton coupling. Similarly, Fig. 5b presents dispersions for select membranes on $SiO_2$ substrates, for which detuning impacts $\Delta$. Combining these results for a chosen $\lambda_p = 400$nm, Fig. 5c compares $\Delta$ from measurements (symbols) against predictions by the Rabi formula (dashed) and numerical calculations (solid curves). Evidently, the simple coupled mode theory underestimates $\Delta$ increasingly for $d < 50$nm compared to both numeric and experimental results, and thus in this regime overestimates the negative group velocity of the antisymmetric mode. This discrepancy originates from the unaccounted dispersion of interface polaritons even before their coupling, owing to their hybridization with free-space light (see SI). Truthfully, the membrane dynamical system couples dispersive interfacial modes $\omega_{SO}^{t,b}(q)$ with frequency-dependent oscillator strengths $\chi_{t,b}(\omega)$ unequal when $\delta\omega > 0$. Thus, when "mode repulsion" is large ($\Delta^2 \sim \omega_{LO}^2 - \omega_{TO}^2$), the rigorous coupled mode theory presents a non-linear eigenvalue equation (see SI). Summarizing these findings at an application-relevant wavelength $\lambda_p = 200$ nm, Fig. 5d presents the numerically solved mode-splitting owing to interfacial coupling ($\Delta$ less the detuning: $\Delta' \equiv \Delta(\varepsilon_{\text{sub}}, d) - \delta\omega(\varepsilon_{\text{sub}})$). Estimating the STO optical phonon scattering rate by $\gamma \approx 30$ cm$^{-1}$,[30] $\Delta' > \gamma$ denotes the simultaneous regime of coherent mSPhPs, resolvable negative dispersion, and polariton strong coupling.[35,36] On the other hand, we denote that regime where the exact splitting $\Delta'$ exceeds by more than $\gamma$ that predicted by the Rabi formula as the "anomalous coupling" regime. Here the Rabi formula strongly underestimates mode-splitting and overestimates mSPhPs velocities, and negative dispersion is essentially dictated by the nonlinear eigenvalue problem (SI). In sum, suspended membranes with thickness much less than the target wavelength $\lambda_p$ sustain polaritons with most *distinct* "negative dispersion", albeit quantitatively underestimated by the Rabi formula.

Thus far, mSPhPs with negative dispersion have been restricted to the range of $\omega_-$ between $\omega_{LO}$ and the larger of $\omega_{SO}^b$ and $\omega_{SO}^t$. Heretofore, this narrow regime has largely concealed asymmetric mSPhPs.[21,37] However, a dielectric superstrate ($\varepsilon_{super} = \varepsilon_{sub}+\delta$) can dramatically reduce $\omega_{SO}^{t}$, thus opening the "window" for $\omega_{-}(q)$. In fact, membrane polaritons possess a unique 'feature of merit' we call "bandwidth-tunability" which is not accessible in hyperbolic media. STO in particular exhibits strong bandwidth tunability capable of drastically improving (~5x) both the bandwidth of the negative dispersion ($\Delta\omega_{-}$) and quality factors through encapsulation of the membrane (see SI). We propose a scheme achieving "spectral overlap" of polaritons with negative and positive dispersion across the boundary of a partially encapsulated membrane (Fig. 5e). For example, Fig. 5g predicts (overlapping) loss functions for encapsulated ($\varepsilon_{sub}$=3, $\delta$=0.5) and suspended regions of the same 50nm STO membrane that can host respective emission and detection of negatively refracted polaritons. We define that range of $\omega$ where the emission and detection regions simultaneously sustain asymmetric (with negative group index, $n_g \equiv n_e^- < 0$) and symmetric ($n_g \equiv n_d^+ > 0$) mSPhPs, respectively, as the "Veselago range". In this regime, the interface between $n_e^-$ and $n_d^+$ regions satisfies the conditions for polariton negative refraction,[7,8,48] enabling focus of a polariton source (*e.g.* in-coupling antenna in Fig. 5e, left) into a perfect image (right). Fig. 5g (top) presents numerical simulations (SI) of polariton energy transmission across the Veselago interface. Fig. 5f predicts that $\delta$>0 breaks perfect antisymmetry of the $n_e^-$ mode (field profile inset to Fig. 5f; details in SI) sufficient to "impedance match" its transmission into the symmetric mode ($n_d^+$). Broadband transmission of polaritons across this Veselago range allows selecting the "magnification ratio" $M(\omega) \equiv q_e/q_d$ of emitted to detected polariton momenta, enabling application control (and even modulation) of the "Veselago image" position and magnification according to $\frac{s_{\mathrm{img}}}{s_{\mathrm{obj}}} = M(\omega)$, where $s_{\mathrm{img,obj}}$ denote distances of the image/object from the Veselago interface. Fig. 5f predicts that, at $M(\omega = 710\ \mathrm{cm}^{-1}) = 1$ (cross in Fig. 5g), relative "image intensity" saturates near 30% for large $\delta$. Additional schemes should be explored to further

i) bridge the polariton "impedance barrier" across our proposed Veselago interface, and ii) improve the quality factor of polaritons resolved in this study (bulk STO supports long-lived polaritons[31]). This latter point is readily resolved in the 2nd STO reststrahlen band (RB2) ($175cm^{-1} - 475cm^{-1}$) where lower intrinsic losses allow for high-quality ($Q_q > 20$) mSPhPs with lifetimes comparable to hyperbolic phonon-polartions in hBN and SPhPs in SiC. While the bandwidth $\Delta\omega_-$ for suspended STO is severely diminished in RB2, the wide range between $\omega_{TO,2}$ and $\omega_{LO,2}$ provides ample room for increasing $\Delta\omega_-$ through encapsulation. A more detailed discussion of quality factor and bandwidth is provided in **Supplementary Discussion 5** alongside comparisons of STO to other common polaritonic platforms.

By harnessing membrane thickness, dielectric environment, structural inhomogeneity, and ultimately composition, we foresee on-demand *transformation optics* with strongly coupled membrane polaritons can enable novel photonics far beyond the Veselago scheme proposed here. Most complex oxides remain unexplored in membrane form, but invite opportunities for multifunctional optics unrealizable in bulk, leveraging facile strain-tuning of electronic and structural phases,[20,49,50] flexoelectric response,[51] and Moiré-patterned ferroicity.[52] For instance, harnessing the local field effect[23] of switchable ferroelectricity in STO membranes[20] could modulate metallicity of 2D adlayers, like graphene. Hetero-layered ferroelectric oxide membranes thus promise *in situ* switchable tuning of unconventional polariton dispersions at the local scale of ferroic domains (<100 nm).[20] Overall, this work establishes free-standing oxide membranes as an extensive platform for transformation optics at the nanoscale.

## Data Availability

All the data that support the findings of this study are reported in the main text and Supplementary Information. Source data are available from the corresponding authors on reasonable request.

## Acknowledgements

B.L., L.T., D.U., H.B., and A.S.M. acknowledge partial support from the Office of the Under Secretary of Defense for Research and Engineering under award number FA9550-22-1-0330. Synthesis of membrane and structural characterization (S.V. and B.J.) were supported by the U.S. Department of Energy through DE-SC0020211, and in part by the Center for Programmable Energy Catalysis, an Energy Frontier Research Center funded by the U.S. Department of Energy, Office of Science, Basic Energy Sciences at the University of Minnesota, under Award No. DE-SC0023464. S.C. acknowledge support from the Air Force Office of Scientific Research (AFOSR) through Grant Nos. FA9550-21-1-0025. Film growth was performed using instrumentation funded by AFOSR DURIP awards FA9550-18-1-0294 and FA9550-23-1-0085. Parts of this work were carried out at the Characterization Facility, University of Minnesota, which receives partial support from the NSF through the MRSEC program under award DMR-2011401. Exfoliation of films and device fabrication was carried out at the Minnesota Nano Center, which is supported by the NSF through the National Nano Coordinated Infrastructure under award ECCS-2025124. S.H.P., and T.L. acknowledge partial support from the Office of Naval Research MURI Grant No. N00014-23-1-2567.

## Author Contributions

B.L, B.J., and A.S.M. conceived the polariton strong coupling experiments, with theory guidance from S.H.P. and T.L. Nano-optics experiments were conducted by B.L. and A.S.M. with instrumental assistance from L.T., D.U., and H.B. SrTiO3 membranes were prepared by hybrid molecular beam epitaxy and transferred to target substrates by S.V., S.C. and B.J. Etched target substrates were prepared by D.S., S.K., and S.H.O. Loss function calculations were performed by B.L., S.H.P., and T.L. Analytic models for interfacial polariton strong coupling were developed by S.H.P. and T.L., who also performed simulations of transmission across the proposed Veselago interface.

## Online Methods:

**Implementation of infrared nano-imaging and -spectroscopy:** Our experiments are enabled by coupling a custom-configured scattering-type scanning near-field optical microscope (s-SNOM) based on a commercial atomic force microscope (NX-10 AFM, *Park Systems*) to a tunable ultrafast infrared laser. The light source combines a 40 MHz pulsed oscillator and optical parametric amplifier (Primus and Alpha, respectively, *Stuttgart Instruments*) supplying near-infrared radiation from an optical parametric amplifier to an integrated difference frequency generation stage producing infrared pulses of spectral bandwidth 50 $cm^{-1}$ (350 fs pulses) centered at energies tunable from 600-2000 $cm^{-1}$.[53] Radiation is monochromated (*Stuttgart Instruments*) to 5 $cm^{-1}$ bandwidth for imaging experiments. This radiation is focused by nano-positioned parabolic mirror onto a sharp (~10nm apex radius) atomic force microscope (AFM) probe (PtSi-FM, *NanoAndMore USA*) whose tip radius determines the spatial resolution of s-SNOM.[17–19] The incident light polarizes the AFM tip which, in turn, polarizes the sample according to the complex, frequency dependent permittivity of the sample volume within the immediate vicinity of the probe.[54] The local polarization response of the sample is contained in probe-scattered light collected by the parabolic mirror and directed into a nitrogen-cooled mercury cadmium telluride photodetector (Judson Teledyne). The photovoltage is demodulated (HF2LI, *Zurich Instruments*) at harmonics of the probe tapping frequency to suppress background-scattered radiation unrelated to the probe-sample near-field interaction. Our imaging (spectroscopy) applies demodulation at the $2^{nd}$ ($3^{rd}$) harmonics of the probe tapping frequency. Simultaneous recording of amplitude and phase of the probe-scatted light (phasor) is enabled by an asymmetric Michelson interferometer with reference mirror alternatively phase-modulated with a piezoelectric actuator for nano-scale imaging (time-resolved pseudo-heterodyne detection

scheme[55]) or scanned by a voice-coil stage (scanDelay 50, *APE*) for nano-scale Fourier transform infrared (nanoFTIR) spectroscopy recorded at 3 $cm^{-1}$ resolution.[19] After normalizing the probe-scattered phasor to that recorded from a non-absorbing medium (e.g. substrate), amplitude and phase are interpreted as reflective and absorptive components ("reflectance" and "absorption") of the surface response, respectively. Unless otherwise stated, all imaging (spectroscopy) results are normalized to the substrate ($SrTiO_3$ membrane interior), respectively. In the latter case "absorption" thus indicates an excess over that recorded from the membrane interior, owing to polariton constructive interference. "Line-scan" nano-spectroscopy is enabled by positioning the AFM probe along a pre-defined path of pixels, typically from a membrane edge to its interior, and at each position recording a probe-scattered spectrum by nano-FTIR. Owing to the limited energy span at a single tuning of our light source, to record a broadband spectrum line-scan, the line-scan path is re-traced several times at incremental tunings of our laser (*e.g.* 5 tunings for Figs. 4b-c). The resulting complex-valued spectra are simply co-added to yield a broadband measurement of nano-reflectance and absorption. This approach leverages the ultra-high long-term positional stability of our AFM system (<50 nm positional drift per hour) and spectral stability of our ultrafast light-source.[53]

**Synthesis and transfer of free-standing $SrTiO_3$ (STO) membranes:** An epitaxial single-crystalline 5 nm SrO sacrificial layer and 10-120 nm thick STO (001) film were grown on a 5 × 5 mm LAO (001) substrate employing a hybrid molecular beam epitaxy (MBE) system (Scienta Omicron Inc). An elemental solid source of Sr (99.99% purity, Sigma Aldrich) sublimated by a thermal effusion cell was used to supply Sr at a beam equivalent pressure (BEP) of $8.4 \times 10^{-8}$ Torr for SrO and STO growth. Meanwhile Ti was supplied using a metal-organic precursor of titanium tetra isopropoxide (TTIP, 99.999%, Sigma Aldrich) injected via gas injector using a vapor inlet system at a BEP of 8.4 x $10^{-6}$ Torr for absorption-controlled stoichiometric STO growth. A radio frequency (RF) plasma source (Oxford Instruments) was employed to supply atomic oxygen species at an oxygen gas pressure of 5 x $10^{-6}$ Torr at 250 W. For STO and SrO growth, the LAO substrate was ramped up to a growth temperature of 950 ℃ (thermocouple temperature) and cleaned by oxygen plasma operating for 25 minutes. After cleaning the substrate via oxygen plasma, an epitaxial SrO layer was grown by co-deposition of Sr and oxygen plasma source for 5 minutes, and subsequently, STO film was grown by co-deposition of Sr, TTIP, and oxygen plasma.

The film surface was monitored in-situ using reflection high-energy electron diffraction (RHEED), Staib Instruments, during the SrO and STO growth. An atomic force microscopy (AFM) was used to characterize the film surface (Nanoscope V Mulitmode 8, Bruker). The structural properties of the film were characterized by high-resolution X-ray diffraction (SmartLab XE, Rigaku) having a Cu K$\alpha$ X-ray source.

STO membranes were transferred onto 10 × 10 mm 300 nm-thick $SiO_2$/Si substrates, bare Si substrates, and pattern-etched Si substrates, using a PDMS stamp in steps: 1) The PDMS stamp was first attached to the STO/ SrO/ LAO sample. 2) The entire PDMS/ STO/ SrO/ LAO sample was placed in water at room temperature for 5 minutes to dissolve the SrO layer. 3) After dissolving the SrO sacrificial layer, the LAO substrate was separated using tweezers. The residual water on the STO membrane was blown out by a $N_2$ gun. 4) Finally, the STO membrane with the PDMS was placed on the target substrate and the PDMS stamp was gently detached from the substrate by tweezers, thus transferring the STO membrane to the target substrate.

**Polariton interferometry and determination of polariton dispersion:** In our polariton imaging s-SNOM experiments, the conductive AFM tip also acts as an optical antenna to both launch and detect phonon-polaritons at the surface of our sample (Fig. 1a). In our case, when tip-scattered photons simultaneously supply energies $\omega = \omega_p$ and momenta (inverse confinement) $q = q_p$ matching those of a polariton resonance at the sample surface, membrane surface phonon-polaritons (mSPhPs) are emitted. Notably, the sharp probe naturally supplies a range of momenta $q \leq 1/a$, with $a \approx 10$nm the probe radius, suitable for "launching" nano-scale polaritons.[39–41] In our case, real-space characteristics of emitted mSPhPs are observable only after they reflect from an edge, defect, or other boundary on the sample, and propagate back towards the AFM tip. Probe-scattered light associates with both the local surface response (e.g. the "emitted" polariton) as well as the nonlocal response from edge-reflected mSPhPs. In imaging, constructive (destructive) interference of electric fields from the emitted and reflected polaritons results in periodic bright (dark) fringes in the probe-scattered absorption response near membrane edges or other polariton reflectors. The polariton half-wavelength is conventionally measurable by recording the distance between bright (or dark) fringes as a function of the energy of incident photons.

When only a single fringe is observable (owing to polariton dissipation), we apply the following simple model, which provides a satisfactory simplification of a more complete model[54] of probe-sample near-field interaction: Representing the $E_z$-field amplitude scattered by the membrane substrate with a phasor $\tilde{e}_0$, and that of emitted polariton with a phasor $\tilde{e}_p$, the total field scattered by the sample at the probe positioned a distance $x$ from the membrane edge is approximately $\tilde{E}_z(x) \approx \tilde{e}_0 + \tilde{e}_p\left(1 - \sqrt{\frac{a}{a+x}}\, e^{2iq_p x}\right)$, with $q_p = 2\pi/\lambda_p$ the (complex) polariton wave-vector, $a$ the probe radius, and $i$ the imaginary unit. The second term in parentheses denotes the reflected ("image") polariton field whose negative sign enforces the requirement $\tilde{E}_z(x=0) - \tilde{e}_0 \approx 0$ associated with vanishing polariton polarization charge when the probe is scanned beyond the membrane. The spatial dependence simplifies that of the polariton Green function, proportional to the radially ($r$) outgoing Hankel function $-iH_0^{(1)}(q_p r)$.[56] In our experiments, normalization to the non-absorptive substrate response renders $\tilde{e}_0$ unity and the polariton phasor $\tilde{e}_p$ entirely imaginary (absorptive). In this model, (1) the absorption response $\mathrm{Im}\{\tilde{E}_z/\tilde{e}_0\}$ is first maximized at $x \approx \lambda_p/2$, and (2) the reflective response $\mathrm{abs}\{\tilde{E}_z/\tilde{e}_0\}$ is first maximized at a shorter distance closer to $x \approx \lambda_p/4$ (cf. Fig. 3b) but depending on the phase $\alpha \equiv \tan^{-1}(|\tilde{e}_p|/|\tilde{e}_0|)$ of the normalized probe-scattered field. Throughout, we deduce the polariton wavelength from the spatially resolved absorption response using (1). In cases where measurement noise renders the "fringe" in probe-scattered amplitude more detectable than in the absorption channel, we deduce the polariton wavelength from (2), whence $\alpha \approx 1$ radians throughout the STO Restrahlen band (see *e.g.* Fig. 2a) yields $\mathrm{abs}\{\tilde{E}_z/\tilde{e}_0\}$ maximized at $x \approx \lambda_p/6$ (*cf.* Fig. 3b). In either case, our measurement of $\lambda_p$ versus probe energy $\omega$ allows to record the polariton dispersion through $q_p(\omega) = 2\pi/\lambda_p(\omega)$ or, equivalently, its inverse relation $\omega_p(q)$.

**Fabrication of patterned (trenched) silicon substrates**: Silicon wafers were cleaned using acetone, isopropanol, and deionized water, followed by a nitrogen blow-dry. A thin layer of polymethyl methacrylate (PMMA 950A, C4) resist was spin-coated onto the substrate at 3000 rpm for 60 seconds. The resist-coated samples were then soft-baked at 180°C for 180 seconds. The desired trench pattern was defined using an electron-beam lithography system (Vitec, EBPG5000+), with an accelerating voltage of 100 kV and a dose of 1,400 μC/cm². After exposure, the samples were developed in a MIBK:IPA (1:3) solution for 60 seconds, followed by an IPA rinse.

The patterned silicon wafers were subjected to reactive ion etching (RIE) using a $CF_4$ + $O_2$ gas mixture to etch the exposed silicon areas. The etching was carried out in a reactive ion etcher (Advanced Vacuum, Vision 320) with a $CF_4$ flow rate of 32 sccm and $O_2$ flow rate of 8 sccm. The chamber pressure was maintained at 80 mTorr, and the RF power was set to 100 W. The etching time was adjusted based on the target trench depth, typically around 25 seconds to achieve trench depths of ~50 nm. The $CF_4$ provided fluorine ions for isotropic etching, while $O_2$ enhanced anisotropy and improved sidewall profile. After the etching process, the PMMA resist was removed by soaking in acetone overnight, and any remaining PMMA was completely removed using oxygen plasma at 100 W for 5 min. The resulting Si trench structures were then analyzed by scanning electron microscopy (SEM) and atomic force microscopy (AFM) to verify the trench dimensions and profile.

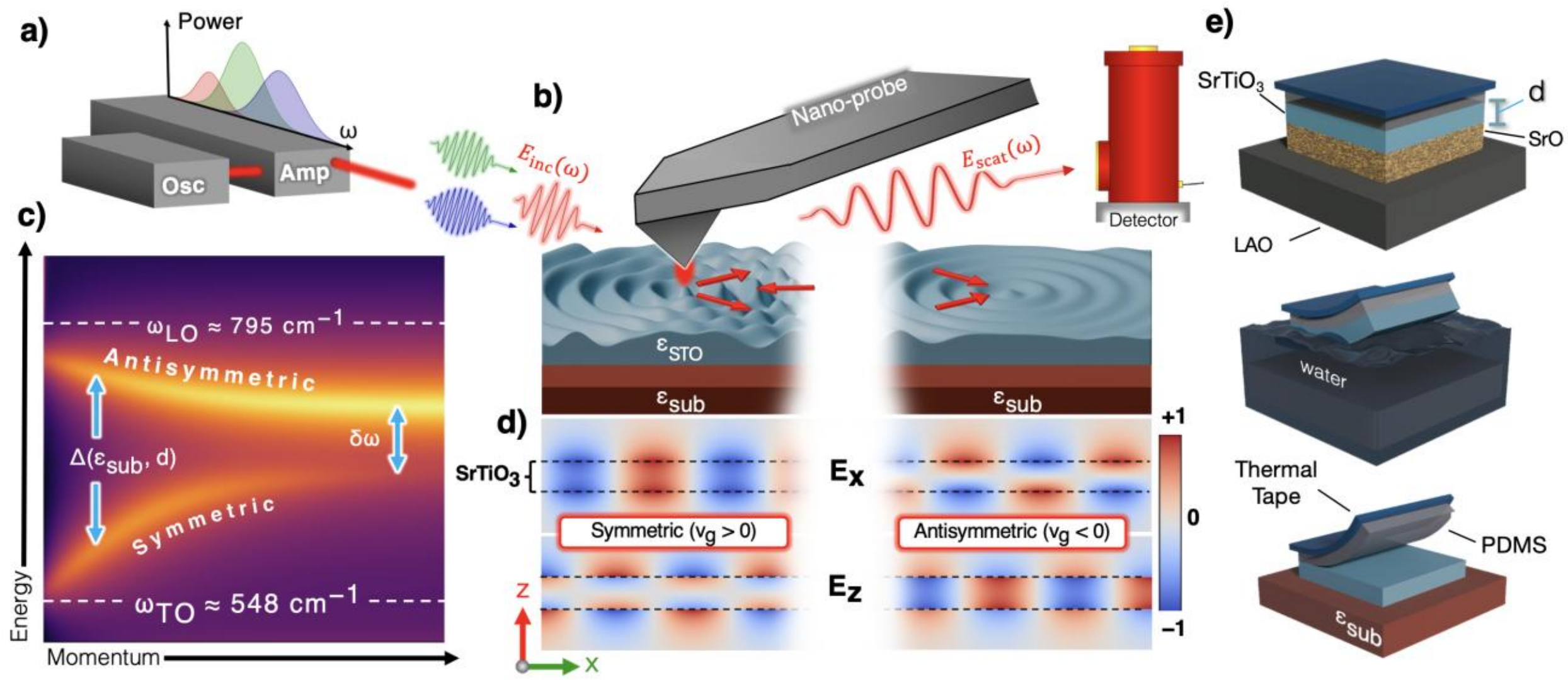

**Figure 1 | Polariton interferometry in free-standing oxide membranes.** **a)** Under illumination field $E_{\mathrm{inc}}(\omega)$ from a tunable ultrafast laser, **b)** scanning a sharp nano-probe over the $SrTiO_3$ membrane surface while recording (at detector) the probe-scattered field $E_{\mathrm{scat}}(\omega)$ enables infrared nano-imaging and -spectroscopy (Methods). Polaritons reflecting from an interface (red arrows) enable polariton interferometry (Methods); those transmitting to a region of negative group velocity are predicted to realize Veselago lensing to an image. **c)** Numerical calculations of the loss function for a 50nm $SrTiO_3$ membrane on an $SiO_2$ substrate. Coupling of interfacial polaritons with detuning $\delta\omega$ results in symmetric ($v_g > 0$) and antisymmetric ($v_g < 0$) normal modes split by $\Delta$. **d)** In-plane (top) and out-of-plane (bottom) electric field profiles of normal modes described in (c). **e)** Free-standing membranes are grown by hybrid molecular beam epitaxy (Methods), separated from their substrate by dissolving a sacrificial layer, and transferred to target substrates ($\varepsilon_{\mathrm{sub}}$).

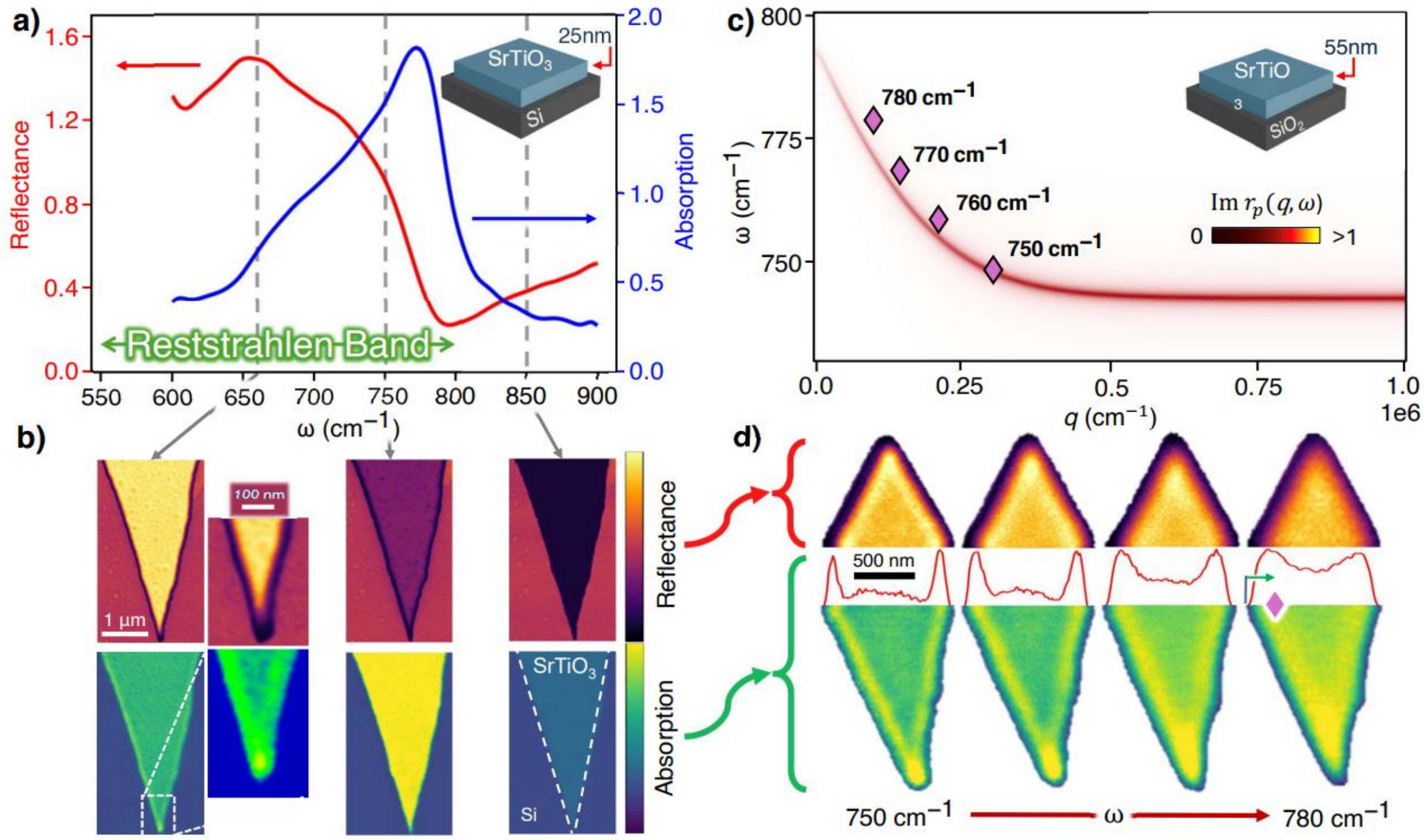

**Figure 2 | Imaging unconventional polariton dispersion in $SrTiO_3$ membranes. a)** Nano-scale infrared spectroscopy of reflectance and absorption (absolute value and imaginary part, respectively, of the probe-scattered field) from the interior a 25nm thick membrane. The gray dashed lines indicate energies of images shown in (b); images and spectra are normalized to the Si substrate. **b)** Near-field reflectivity and absorption acquired at monochromated energies 666cm$^{-1}$, 750cm$^{-1}$, and 850cm$^{-1}$ (left to right). Zoom-in to the membrane's triangular pinnacle reveals polariton interference. **c)** Polariton momenta $q_p(\omega)$ extracted from wavelengths recorded in (d) trace a dispersion matching the loss function predicted for the antisymmetric mode of this membrane. **d)** Monochromated images of mSPhPs within a 1 micron-wide diamond-shaped membrane flake of 50nm thickness. Dispersing fringes are present near the membrane edge in both reflectivity (top) and absorption (bottom) images; line-cuts (red) show their wavelength increases with frequency.

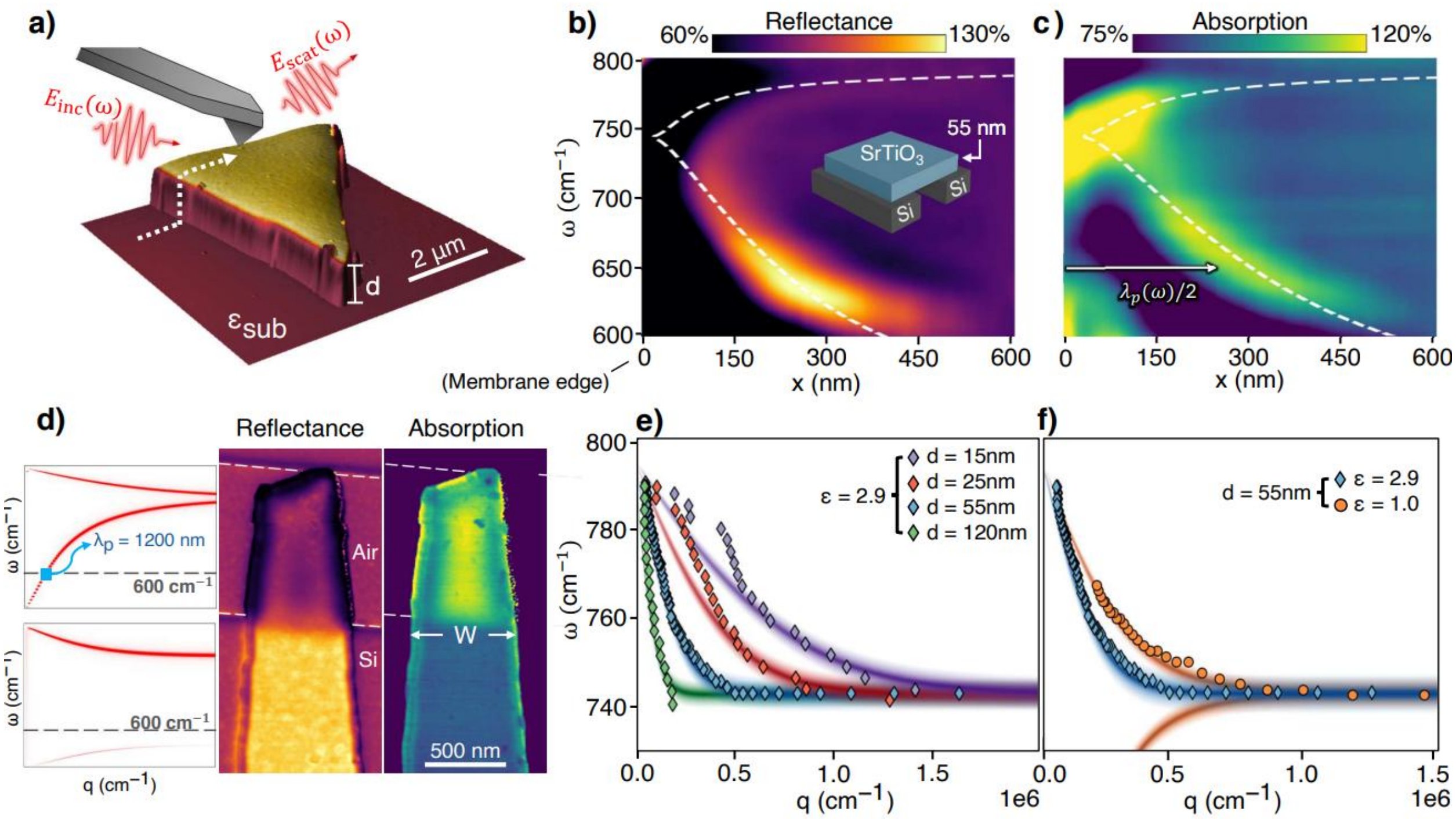

**Figure 3 | Substrate and thickness tuning of negative dispersion. a)** Line-scan nano-spectroscopy is realized by recording spectra of probe-scattered radiation at coordinates near the membrane edge. **b-c)** Nano-spectroscopy from a 55nm-thick suspended $SrTiO_3$ membrane (inset). The dashed lines indicate the real-space dispersions predicted for polariton amplitude and phase (Methods). For visibility of polariton fringes, spectra are presented normalized to the spectral response of the membrane interior. **d)** Reflectance and absorption imaging of a membrane (width *W*) partially suspended over a 90nm deep trench in Si. At 600cm-1, the loss function (left; axes shared with Fig. 1c) predicts $\lambda_p=2W$ in the suspended membrane (top) and no mSPhPs in the supported membrane (bottom); color scales shared with b-c). **e-f)** Nano-spectroscopy from several membranes reveals the thickness- (e) and substrate-dependent (f) tunability of the mSPhP negative dispersion; associated loss functions are superimposed for brevity (positive dispersion in (f) associates with the suspended membrane, $\varepsilon = 1$).

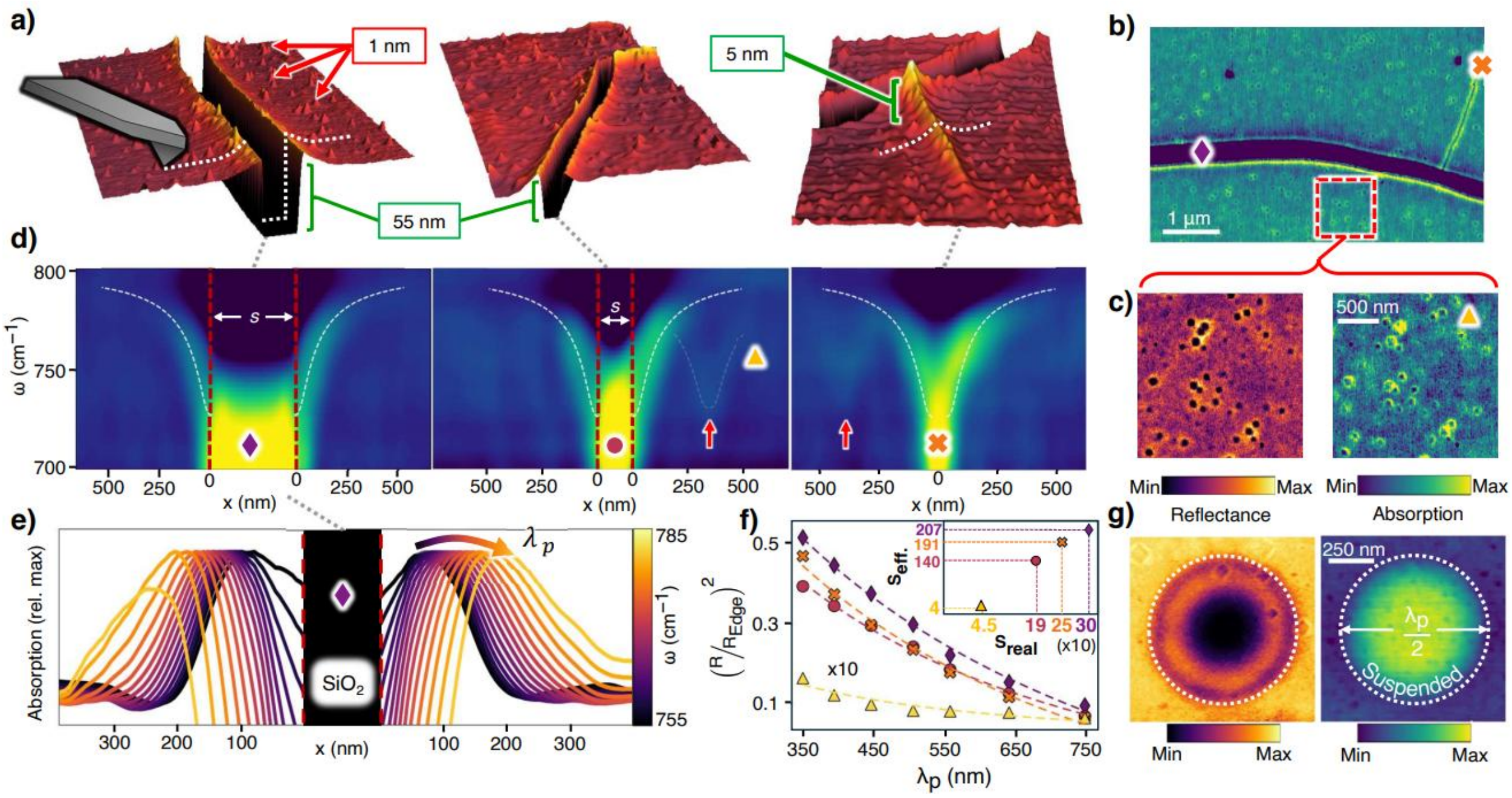

**Figure 4 | Defect-scattering and confinement of membrane polaritons. a)** AFM topography of defect structures in a 55nm $SrTiO_3$ membrane transferred to $SiO_2$. **b-c)** Near-field imaging (750 cm-1) and **d)** line-scan spectroscopy (absorption) reveal mSPhP reflections from these defects; symbols indicate regions in b-c). The dashed lines in (d) indicate predicted dispersions. Red arrows indicate weak polariton scattering from single point defects in (c). **e)** Linecuts from nano-spectroscopy of the large crack reveal the negative dispersion of $\lambda_p(\omega)$; amplitudes of the mSPhP absorption are normalized for easier comparison. **f)** mSPhP reflection as a function of $\lambda_p$ relative to reflections at the membrane edge. Symbols associate with defects in b-d). **(Inset)** Defect sizes $s_{eff.}$ derived from fitting to a simple model of polariton reflection scale with physical sizes identified from (a). **g)** Near-field imaging (600 $cm^{-1}$) of symmetric polaritons confined in a membrane suspended over a circular hole in the substrate with diameter $\lambda_p/2$.

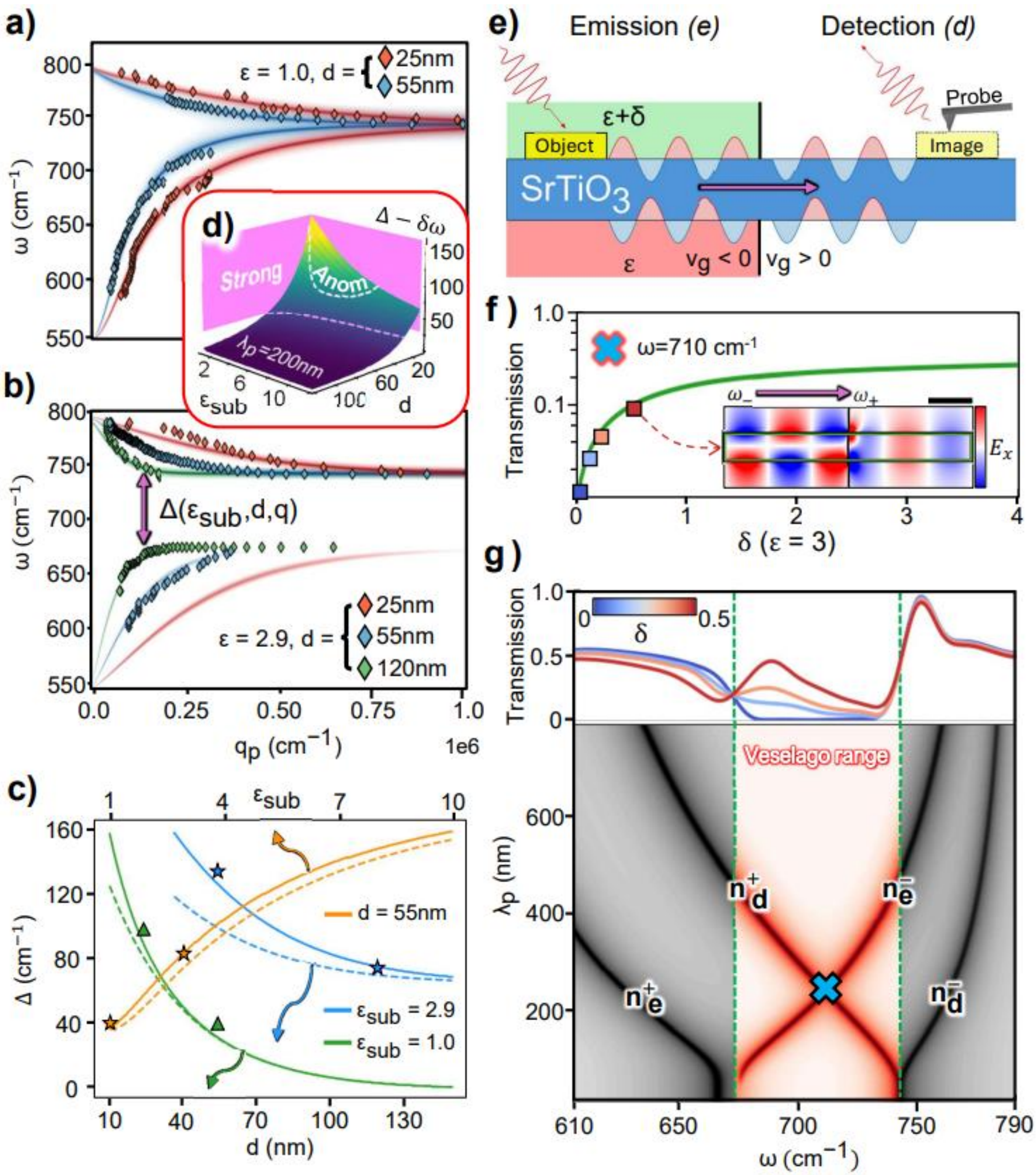

**Figure 5 | Engineering Veselago nano-optics in freestanding membranes. a-b)** Measured dispersions reveal the mSPhP mode splitting Δ in suspended (a) and supported (b) membranes. **c)** Measured (symbols) and predicted (curves) mode splitting versus membrane thickness (blue/green, at $\lambda_p$ = 400nm) and versus substrate permittivity (orange, at $\lambda_p$ = 200nm). Solid and dashed curves compare predictions from the nonlinear eigenvalue problem (SI) and Rabi formula, respectively. **d)** Exemplary thickness and substrate dependence of the mSPhP mode-splitting (Δ less the detuning $\delta\omega$) at $\lambda_p = 200$nm, indicating regimes of polariton strong coupling and

anomalous coupling. **e)** Proposed realization of in-plane Veselago lensing from an object (polariton emission) to an image (polariton detection) in a free-standing membrane. **f)** Simulated transmission efficiency of mSPhPs from the emitted asymmetric polariton mode $E_e^-$ to the detectable symmetric mode $E_e^+$ across the Veselago interface in (e) versus of δ for ω = 710cm$^{-1}$ (at cross in (g), bottom). **(Inset)** Transmitted $E_x$ field from antisymmetric to symmetric modes at ε = 3, δ = 0.5. The scale bar indicates $\lambda_p/2 = 125\text{nm}$ **g)** Simulated transmission spectra (top) and real-space dispersions (bottom) of 50nm $SrTiO_3$ encapsulated by ε = 3, δ = 0.5 (group index $n_e$) and ε = 1 ($n_d$), exhibiting overlapping positive and negative mSPhP group velocity indices ($n_{e,d}^{\pm}$); the Veselago range (red) indicates that span of ω supporting negative refraction.

**Supplementary Information: Interfacial strong coupling and negative dispersion of propagating polaritons in freestanding oxide membranes**

**Authors:** Brayden Lukaskawcez[1], Shivasheesh Varshney[2], Sooho Choo[2], Sang Hyun Park[3], Dongjea Seo[3], Liam Thompson[1], Nitzan Hirshberg[1], Madison Garber[1], Devon Uram[1], Steven Koester[3], Sang-Hyun Oh[3], Tony Low[3], Bharat Jalan[2], Alexander McLeod[1,*]

4. School of Physics and Astronomy, University of Minnesota, Twin Cities, Minneapolis, Minnesota 55455, USA
5. Department of Chemical Engineering and Materials Science, University of Minnesota, Twin Cities, Minneapolis, Minnesota 55455, USA
6. Department of Electrical and Computer Engineering, University of Minnesota, Minneapolis, Minnesota 55455, USA

** Corresponding author:* mcleoda@umn.edu

## Table of Contents

# Supplementary Discussion 1: Influence of residues from membrane transfer

We apply a polymer-based method to transfer $SrTiO_3$ (STO) membranes onto target substrates, which has the possibility to leave polymer residues on the membrane surface. Here we discuss the potential impact of these residues on propagation of polaritons in our membranes. As an exemplary membrane, we consider the 25nm membrane shown in Fig. 2a,b (main text). In addition to the original synthesis and transfer, was transferred by hand a second time, via PDMS and thermal release tape, from its original host substrate to a new substrate leaving behind a non-uniform layer of polymer residue on the surface. This polymer layer appears most clearly in the first panel of Fig. 2b (main text), resembling a bubble-like texture on the surface of the membrane. While we cannot precisely quantify the contribution of the polymer layer to the collected near-field signal, previous topography measurements of the membrane on its original host substrate suggest that the thickness of the polymer is on the order of a few nanometers and therefore is unlikely to contribute significantly to the overall near-field signal, and therefore has no discernible impact on the polariton properties of this membrane nor on the other membranes explored in this work. This conclusion is consistent with former nano-infrared spectroscopy of these membranes, which uncovered evidence for not more than a few nanometers of polymer residue at the interface of comparable STO membranes.[1]

# Supplementary Discussion 2: Accessing membrane polaritons at large detuning

Despite performing line-scans on our 25nm-thick STO membrane on $SiO_2$, we do not plot the symmetric membrane surface phonon polariton (mSPhP) dispersion in Fig. 5b (main text). Loss function calculations predict that the oscillator strength of the symmetric mSPhP decreases with increasing asymmetry of the surrounding environment. Combining this with the overall reduction in nano-optical signal detectable from thinner samples, it is possible that the symmetric mSPhP dispersion is not accessible in our setup for thin membranes at large detuning comparable to that of $SiO_2$. For substrates with infrared permittivity significantly larger than $SiO_2$, such as silicon, the detuning $\delta\omega$ (see main text for definition) becomes so large that $\omega_{SO}^{b}$, the surface optical phonon frequency for the $SrTiO_3$-silicon interface, falls below the lower limit of energy addressable with our combination of laser and detector ($\omega_{sp} < 600$ $cm^{-1}$). For this reason, we are unable to measure the symmetric mSPhP dispersion in any of our samples on silicon substrates.

# Supplementary Discussion 3: Coupled Mode Model and the Rabi formula

We introduce a simple coupled mode model to describe the behavior of mSPhPs in STO membranes. Consider a single planar interface between a material with dielectric constant $\epsilon_1$ and STO with dielectric function $\epsilon_{STO}(\omega)$. In terms of polariton momentum $q$, the dispersion relation of a mSPhP confined to the interface is given by $\frac{\epsilon_1}{\kappa_1(q,\omega)} + \frac{\epsilon_{STO}}{\kappa_{STO}(q,\omega)} = 0$ where $\kappa_j = \sqrt{q^2 - \epsilon_j\omega^2/c^2}$ and the decay length of the mSPhP mode is $\sim 1/\kappa_j$. Modeling STO as a collection of Lorentzian oscillators, we later supply a fully analytic solution for the mSPhP frequency $\omega_{sp,j}$ in the limit $q \to \infty$ given by $\epsilon_{STO}(\omega_{sp,j}) = -\epsilon_j(\omega_{sp,j})$. An STO membrane with thickness d surrounded by dielectrics with permittivities $\epsilon_1$ and $\epsilon_2$ will host two mSPhP modes, one per each STO-dielectric interface. In general, for $\epsilon_1 \neq \epsilon_2$, the frequencies of the mSPhP modes $\omega_{sp,1}, \omega_{sp,2}$ will be different. (The main text associated these with energies $\omega_{SO}^{b,t}$ of interfacial polaritons at bottom and top membrane interfaces.) For sufficiently thin films where $d \sim 1/\kappa_{STO}$ the mSPhP on the two interfaces are expected to couple through their exponentially decaying electric fields. At a given value of q, assuming each SPhP mode to be a resonator with frequency $\omega_{sp,j}$, the total electric field of the slab can be written as a superposition of the independent SPhP modes localized to each interface,

$$E = a_1(t)E_1 + a_2(t)E_2$$

where $E_j = e_j(z)e^{iqx}e^{-i\omega_j t}$ and $e_j(z)$ are the polarization vectors for the electric and magnetic fields. The amplitudes $A(t), B(t)$ describe the time-dependent superposition of the SPhP modes. The power transferred from mode 1 to mode 2 can be written as

$$\int a^*{}_2 E_2{}^* \cdot (-i\omega_1 P_{12}) dz + c.c. = i\omega a_2{}^* a_1 \int \epsilon_0 \Delta\epsilon E_1 \cdot E_2{}^* dz + c.c.$$

where $-i\omega_1 P_{12} = -i\omega_1 \Delta\epsilon E_1$ is the polarization current induced at interface 2 by mode 1 and $\Delta\epsilon = (\epsilon_2 - \epsilon_{STO})\theta(r \in \Omega_2) \geq 0$ is the perturbation required to introduce mode 2, where $\theta$ is nonzero only in volumetric domain $\Omega_2$ where $\epsilon = \epsilon_2$. From coupled mode theory, the power transferred is given by $d|a_2|^2/dt = \kappa_{21} a_1 a_2{}^* + c.c.$. Hence, identifying with the above relation and normalizing with respect to the mode energy gives the coupling coefficient

$$\kappa_{21} = i\omega \frac{\int \Delta\epsilon\, E_1 \cdot E_2{}^* dz}{\int \epsilon\, E_1 \cdot E_1{}^* dz}.$$

Here $\epsilon$ denotes the spatial permittivity in absence of the "perturbation" of $\epsilon_2$, whence $\kappa_{21} < 0$ since $\epsilon_{STO} < 0$ at energies under consideration here. Since the fields of mode 1 and 2 exponentially decay with a factor of $e^{-qd}$ we may write $\kappa_{21} \sim \Delta_0 e^{-qd} < 0$ where $|\Delta_0| > 0$ is a coupling strength that will be determined by the oscillator strengths of the two modes. The eigenmodes of the thin film can be described using a coupled mode model[2] written as

$$\begin{pmatrix} \omega_1 & \Delta_0 e^{-qd} \\ \Delta_0 e^{-qd} & \omega_2 \end{pmatrix} \begin{pmatrix} a_1 \\ a_2 \end{pmatrix} = \omega \begin{pmatrix} a_1 \\ a_2 \end{pmatrix}$$

The eigenfrequencies of the model are

$$\omega_{\mp} = \frac{\omega_{sp,1} + \omega_{sp,2}}{2} \pm \frac{1}{2}\sqrt{\left(\omega_{sp,1} - \omega_{sp,2}\right)^2 + 4\Delta_0^2 e^{-2qd}}.$$

and the mode splitting $\omega_0 - \omega_+$ is given by the Rabi formula:

$$\Delta = \sqrt{\left(\omega_{sp,1} - \omega_{sp,2}\right)^2 + 4\Delta_0^2 e^{-2qd}}.$$

Note that signs on $\omega_{\mp}$ convey the symmetry of the eigenmodes $\psi_{\mp} = (a_1, a_2)_{\mp}$. Owing to $\kappa_{21} < 0$, the larger of $\omega_{\mp}$ corresponds to an antisymmetric eigenmode. This feature is unique to the present geometry and qualitatively corresponds to coupled mode phenomena studied in "insulator-metal-insulator" waveguides.[3] In this phenomenological theory, we fix the coupling magnitude $|\Delta_0|$ to fixed to 90cm$^{-1}$ by fitting to mode splitting identified in the numerically calculated loss function in the weak coupling regime. As summarized in the main text, the quantity $\Delta_0^2 e^{-2qd}$ can be identified with $\Omega(q)^2$, the square of the Rabi frequency, equal to the rate (and oscillation frequency) of coherent energy transfer between interfacial modes $\omega_{1,2}$ in the case where either is selectively populated.[4] (It should be emphasized that this model is approximate and discards inter-mode coupling at equally physical conjugate negative frequencies $-\omega_{1,2}^*$.[5] Nevertheless, this omission does not qualitatively affect the phenomenology of our coupled modes.) All other quantities in the expression are determined analytically from material parameters and experimental conditions.

## Supplementary Discussion 3.1: Calculations of the surface phonon frequency

In the limit $q \rightarrow \infty$, the phonon polariton frequency at the interface between a dielectric $\epsilon_1$ and STO is given by $\epsilon_1 + \epsilon_{STO}(\omega_{sp}) = 0$. The dielectric function of STO is given by

$$\epsilon(\omega) = \epsilon_\infty \prod_{i=1}^{3} \frac{\omega_{LO,i}^2 - \omega^2 - i\gamma\omega}{\omega_{TO,i}^2 - \omega^2 - i\gamma\omega}$$

where the TO frequencies are $\omega_{TO,1} = 90\text{cm}^{-1}, \omega_{TO,2} = 175\text{cm}^{-1}, \omega_{TO,3} = 544\text{cm}^{-1}$, the LO frequencies are $\omega_{LO,1} = 172\text{cm}^{-1}, \omega_{LO,2} = 475\text{cm}^{-1}, \omega_{LO,3} = 796\text{cm}^{-1}$, and $\epsilon_\infty = 5.2$. Assuming zero loss ($\gamma = 0$), the surface phonon frequency is analytically given by

$$\omega_{sp} = \sqrt{-\frac{1}{3A}\left(B + \xi\beta + \frac{\alpha_0}{\xi\beta}\right)}$$

where

$$A = \epsilon_1 + \epsilon_\infty$$

$$B = -\epsilon_1 \sum_i \omega_{TO,i}^2 - \epsilon_\infty \sum_i \omega_{LO,i}^2$$

$$C = \epsilon_1\left(\omega_{TO,1}^2\omega_{TO,2}^2 + \omega_{TO,2}^2\omega_{TO,3}^2 + \omega_{TO,3}^2\omega_{TO,1}^2\right) + \epsilon_\infty\left(\omega_{LO,1}^2\omega_{LO,2}^2 + \omega_{LO,2}^2\omega_{LO,3}^2 + \omega_{LO,3}^2\omega_{LO,1}^2\right)$$

$$D = -\epsilon_1 \prod_i \omega_{TO,i}^2 - \epsilon_\infty \prod_i \omega_{LO,i}^2$$

$$\alpha_0 = B^2 - 3AC$$

$$\alpha_1 = 2B^3 - 9ABC + 27A^2D$$

$$\beta = \left(\frac{\alpha_1 - \sqrt{\alpha_1^2 - 4\alpha_0^3}}{2}\right)^{1/3}$$

$$\xi = \frac{-1 + i\sqrt{3}}{2}$$

## Supplementary Discussion 3.2: Loss function of STO membranes and oscillator strengths of interfacial polaritons

In the following way, the squared rabi frequency $\Omega(q)^2$ can be identified with a product of interfacial polariton oscillator strengths, multiplied by the momentum-dependent form factor $e^{-2qd}$. The momentum ($q$)-dependent Fresnel reflection coefficient $r_p(q)$ for *p*-polarized light for a thickness-*d* STO membrane with semi-infinite sub- and super-strates of permittivities $\varepsilon_b$ and $\varepsilon_t$, respectively, is given in terms of the *p*-polarized reflection coefficients for top and bottom interfaces $r_{t,b}$ as follows:[6]

$$r_p(q) = \frac{r_t(q,\omega) + r_b(q,\omega)e^{-2\kappa_{STO}(q,\omega)d}}{1 + r_t(q,\omega)r_b(q,\omega)e^{-2\kappa_{STO}(q,\omega)d}}.$$

Here again $\kappa_j(q,\omega) = \sqrt{q^2 - \epsilon_j\omega^2/c^2}$ denotes the propagation constant of light within medium $j = t, b, \mathrm{STO}$. The loss function for the membrane is given by Im $r_p(q,\omega)$, whose intensity maxima (*e.g.* Fig. 1c, main text) identify poles of $r_p(q)$ (dispersion of coupled normal modes).

The energy and oscillator strength of interfacial polaritons at top and bottom interfaces can be inferred from the poles of the reflectivities $r_{t,b}$, respectively. These are "ingredients" to normal modes of the fully interacting system, the mSPhPs, which are the poles of $r_p(q)$ . First, at a single interface of the STO, the reflectivity is given by

$$r_{t,b}(q,\omega) = \pm\frac{\epsilon_{STO}(\omega)\kappa_{t,b}(q,\omega) - \epsilon_{t,b}\kappa_{STO}(q,\omega)}{\epsilon_{STO}(\omega)\kappa_{t,b}(q,\omega) + \epsilon_{t,b}\kappa_{STO}(q,\omega)} \equiv \pm\frac{\chi_{t,b}(q,\omega)}{\omega - \omega_{sp}^{t,b}(q)}$$

where we have made the existence of the interfacial surface phonon polaritons (poles) at $\omega_{sp}^{t,b}$ explicit. The phonon polariton loss can be introduced by $\omega_{sp} \to \omega_{sp} + i\gamma$, where $\gamma \approx 21$ cm$^{-1}$ quantifies the typical scattering rate for the optical phonon in crystalline STO, which is representative for our membranes.[1,7] Thus, evaluating $r_{t,b}$ in proximity to $\omega \approx \omega_{sp}^{t,b}$ we find the oscillator strengths $\chi_{t,b}(q)$ (oscillator residues) of interfacial polaritons can be written as

$$\chi_{t,b}(q,\omega) = i\gamma\frac{\epsilon_{STO}\kappa_{t,b}(q) - \epsilon_1\kappa_{t,b}(q)}{\epsilon_{STO}\kappa_{t,b}(q) + \epsilon_{t,b}\kappa_{t,b}(q)}.$$

Note that all quantities in this expression are evaluated at $\omega \approx \omega_{sp}^{t,b}$, although perfect equality to $r_{t,b}(q,\omega)$ requires broader evaluation at arbitrary $\omega$ within the STO Restrahlen band. Even in the simplest approximation where $\omega = \omega_{sp}^{t,b}(q)$, the finite $q$-dispersion of interfacial polaritons endows the oscillator strengths with intrinsic $q$-dependence that is inaccessible to the simple coupled mode model presented earlier.

## Supplementary Discussion 3.3: Nonlinear eigenvalue problem of coupled membrane surface phonon polaritons

The dispersions of mSPhPs are given by the poles $\omega_{\mp}(q)$ of $r_p(q)$ solving the equation:

$$(\omega_{\mp} - \omega_{\mathrm{sp}}^{t}(q))(\omega_{\mp} - \omega_{\mathrm{sp}}^{b}(q)) - \chi_{\mathrm{t}}(q,\omega_{\mp})\chi_{\mathrm{b}}(q,\omega_{\mp})\,\mathrm{e}^{-2\kappa_{\mathrm{STO}}(q,\omega_{\mp})d} = 0.$$

These correspond with normal modes of the nonlinear dynamical system and with solutions $\omega = \omega_{\mp}$ of a *nonlinear eigenvalue problem*:

$$\begin{pmatrix} \omega_{\mathrm{sp}}^{t}(\mathrm{q}) & g(q,\omega) \\ g(q,\omega) & \omega_{\mathrm{sp}}^{b}(\mathrm{q}) \end{pmatrix} \begin{pmatrix} a_1 \\ a_2 \end{pmatrix} = \omega \begin{pmatrix} a_1 \\ a_2 \end{pmatrix}.$$

Here the energy-dependent coupling constant is given by:

$$g(q,\omega) \equiv -\sqrt{\chi_{\mathrm{t}}(q,\omega)\chi_{\mathrm{b}}(q,\omega)}\,\mathrm{e}^{-\kappa_{\mathrm{STO}}(\mathrm{q},\omega)d}.$$

For a chosen momentum $q$, the case where solutions $\omega$ differ weakly from $\omega_{\mathrm{sp}}^{t,b}(q)$ allows to evaluate $\chi_{t,b}$ at $\omega \approx \omega_{\mathrm{sp}}^{t,b}(q)$ where the nonlinearity is irrelevant. This case reproduces the linear coupled mode theory presented earlier and corresponds to weak coupling scenarios where $\kappa_{\mathrm{STO}} d \approx qd \gg 1$ and/or $\delta\omega \equiv \left|\omega_{\mathrm{sp}}^{t} - \omega_{\mathrm{sp}}^{b}\right| \gg \sqrt{\chi_{\mathrm{t}}(q,\omega)\chi_{\mathrm{b}}(q,\omega)}$. Thus, in the strong coupling limit where the (dynamically determined) squared Rabi frequency satisfies $\Omega^2(q,\omega_{\mp}) = 4\chi_{\mathrm{t}}(q,\omega_{\mp})\chi_{\mathrm{b}}(q,\omega_{\mp})e^{-2qd} > \gamma^2$, we expect that quantitative predictions of coupled normal mode dispersion must address the nonlinearity of the underlying eigenvalue problem.

Our discussion in the main text compares measurements of $\Delta(q)$ with predictions by i) the linear coupled mode model (the Rabi formula) and ii) numerical calculations of the loss function. Although the latter method faithfully captures physics relevant to the nonlinear dynamical system, is it a retrospective approach, and not a deterministic tool for "dispersion engineering". Therefore, direct solution to the nonlinear should provide improved guidance for harnessing membrane polaritons and similar dynamical photonic systems for transformation optics at the nanoscale.

# Supplementary Discussion 4: Simulations of Veselago transmission

The finite element solver Comsol Multiphysics was used to compute i) normal mode field profiles presented in Figs. 1 and 5 of the main text, ii) polariton transmission across the Veselago interface presented in Fig. 5f-g of the main text and iii) real-space propagation patterns of polaritons in an in-plane Veselago lens presented in Fig. S8. These simulations were conducted according to the following configuration. A schematic representation of a 2D cut along the *xz*-plane of the simulation domain is shown in Fig. S1. T structure extends infinitely in the *y*-direction. The STO membrane is arrayed in the *xy*-plane. Perfectly matched layers (not shown in figure) are added around the simulation domain. STO is modeled as a dielectric with thickness $\mathrm{d_{STO}}$ and dielectric function $\epsilon_{\mathrm{STO}}(\omega)$ and extends through the entire simulation domain. The dielectric environment is separated into two domains. The right domain (see Fig. S1) represents a suspended STO membrane. The left domain is encapsulated by dielectrics with dielectric constants $\epsilon$ and $\epsilon + \delta$. The $\delta$ parameter breaks the symmetry of the enviroment and enables transmission from an antisymmetric mode into a symmetric mode. The phonon polariton mode is launched using a z-oriented dipole moment placed 5nm above the STO membrane (red arrow in Fig. S1). Transmission of the phonon mode is quantified as the ratio of the $\mathrm{H_y}$ field magnitude at the STO surface in the two domains. As noted above, it is essential to break the symmetry of the dielectric environment such that the modes are no longer perfectly symmetric or anti-symmetric. The effect of symmetry-breaking on the dispersion and field profiles is shown in Fig. S2. The modes are now a mixture of symmetric and anti-symmetric components which allows transmission into the suspended region that supports symmetric and anti-symmetric modes. According to our real-space simulations, Veselago foci are readily identified provided that the group quality factor (**Supplementary Discussion 5**) exceeds $Q_G = 3$ or the source-interface distance does not exceed 1 polariton wavelength.

# Supplementary Discussion 5: Quality factor in the strong coupling description

Here we calculate the quality factor for mSPhPs using the strong coupling description. Consider the roots solving the first equation of **Supplementary Discussion 3.3,** in which selection of real (complex) momenta $q$ obtains complex (real) values for the mSPhP frequencies $\omega_{\pm}$. The traditional approach to calculating the quality factor for such polariton modes is to take $\omega_{\pm} \in \mathbb{R}$ whereby accompanying values of $q$ are complex. Here a quality factor for polariton propagation can be defined as $Q_q = \left|\frac{Re(q)}{Im(q)}\right|$, representing physically the number of wavelengths the polariton can propagate spatially before decaying to $e^{-1}$ (with $e$ the base of the natural logarithm) of its original amplitude. On the other hand, in our strong coupling description, taking $q$ to be real and $\omega_{\pm}$ to be complex-valued yields a quality factor $Q_\omega = \left|\frac{Re(\omega)}{Im(\omega)}\right|$ that represents the number of temporal oscillation periods the polariton sustains before decaying to $e^{-1}$ of its original amplitude. Moreover, when this description is generalized to any damped oscillator membrane permittivity $\epsilon(\omega)$ ($\approx \epsilon_{\mathrm{STO}}(\omega)$ in our case), the nonlinear eigenvalue problem described in **Supplementary Discussion 3.3** is analytically tractable with $\omega_{\pm}$ obtained as selected complex roots of a third order polynomial for fixed (real) $q$. Recalling our definition of the (complex) group index of refraction $\left(n_g \equiv \frac{v_g}{c} = \frac{1}{c}\frac{\partial \omega_{\pm}}{\partial q}\right)$, the real and imaginary parts of $n_g$ describe the rate of energy propagation and dissipation, respectively. In an analogous way to the FOM for negative-index plasmon waveguides, we define the group quality factor in terms of decaying energy density $Q_G = \frac{1}{2}\frac{Re(n_g)}{Im(n_g)}$, where the additional factor of ½ accounts for the squared field amplitude dependence of the energy density. Consistent with previous studies[ref. Hillenbrand], we will take this approach to assess mSPhP quality factors for a variety of candidate polariton media in comparison to $SrTiO_3$. We first deduce this polynomial, then discuss its solutions in the context of various conceivable membrane polariton materials when treated with appropriate oscillator permittivities.

Considering for simplicity a symmetric dielectric environment ($\epsilon_t = \epsilon_b = \epsilon_0$), we can rearrange our governing self-consistent equation for the mSPhP dispersion in terms of effective (and not necessarily equal) Rabi splittings $\Omega_{\mp}$ as follows:

$$\Omega_{\mp}/2 = (\omega_{\mp} - \omega_{sp}(q)) = \pm i\gamma\left(\frac{\epsilon_{\mathrm{STO}} - \epsilon_0}{\epsilon_{\mathrm{STO}} + \epsilon_0}\right)e^{-qd}.$$

Here the $\mp$ corresponds respectively to antisymmetric/symmetric mSPhP modes, and we applied $\kappa \approx q$ in the limit $q \gg \omega/c$. We now treat the membrane permittivity $\epsilon_{\text{STO}} = \epsilon_\infty \left(1 + \frac{A^2}{\overline{\omega}_0^2 - \overline{\omega}^2 - i\overline{\omega}}\right)$ as a single Lorentz oscillator in terms of the dimensionless frequency $\overline{\omega} \equiv \omega/\gamma$, with dimensionless transverse mode frequency $\overline{\omega}_0 \equiv \omega_{TO}/\gamma$ and oscillator strength $A^2 \equiv \left(\frac{\omega_{LO}}{\gamma}\right)^2 - \overline{\omega}_0^2$. Her we have selected the mode damping $\gamma$ as the fundamental energy scale, with $\overline{\omega}_0 > 0$ effecting merely an energy up-shift of the Restrahlen band and consequent polariton activity. This parameterization generalizes to candidate membrane polariton media that showcase a single dominant optical phonon (e.g. oxides or ionic crystals) or even a Drude response (*i.e.* metals, whereby $\overline{\omega}_0 = 0$ and $\omega_{LO}$ signifies the plasma frequency). For fixed real-valued polariton momentum $q$ relative to the membrane thickness $d$, the complex mSPhP frequencies $\omega_\mp$ can now be expressed as the roots of a cubic polynomial in $\overline{\omega}_\mp \equiv \frac{\omega_\mp}{\gamma}$:

$$(\overline{\omega}_\mp - \overline{\omega}_{sp})\left((\epsilon_\infty + \epsilon_0)(\overline{\omega}_0^2 - \overline{\omega}_\mp^2 - i\overline{\omega}_\mp) + \epsilon_\infty A^2\right) \mp i\left((\epsilon_\infty - \epsilon_0)(\overline{\omega}_0^2 - \overline{\omega}_\mp^2 - i\overline{\omega}_\mp) + \epsilon_\infty A^2\right) e^{-qd} = 0.$$

Membrane polaritons are identified wherever this polynomial admits a root in the fourth quadrant of the complex $\omega$ plane, and the negatively dispersing antisymmetric polariton is selected with (-) in the above equation.

**Supplementary Table 2** presents an enumeration of popular polaritonic media and their single-oscillator parameterizations within this framework. For each such candidate, we compute quality factors $Q_\omega$ and $Q_G$ according to the materials' oscillator parameters. Applying our analysis to bulk STO we find $Q_\omega \approx 560$ within the 2nd Reststrahlen band (RB2) and $Q_\omega \approx 177$ within the 3rd Reststrahlen bands (RB2 and RB3), respectively. This is contrasted with recent experimental determinations of polariton quality factors in STO membranes which found $Q_q \approx 22$ (RB2) and $Q_q \approx 6$ (RB3). We attribute this disagreement largely to variations in optical properties between bulk and thin-film STO [1]. Nevertheless, it is worth noting that we do find agreement between experiment and theory for relative quality factors computed using our simple single-oscillator model. For example, comparing the ratios $\frac{Q_q}{Q_{q,RB3}}$ and $\frac{Q_\omega}{Q_{\omega,RB3}}$ one finds approximate agreement for many of the materials in **Supplementary Table 2** such as SiC ($Q_q \approx 28$)[8], hBN ($Q_q \approx 20$)[9], as well as Ag-based waveguides ($Q_q \sim 4$)[3].

In fig. S6d. we plot the group quality factor $Q_G$ for the $\overline{\omega}_-$ mode as a function of $A$ and $\overline{\omega}_0$. For simple comparison of polariton fitness, we plot names of several polaritonic media according to their oscillator parameters ($A$, $\overline{\omega}_0$). Focusing on STO, RB3 exhibits small $Q_G$ comparable to that of AlN membranes and Ag waveguides. In contrast, RB2 boasts impressive group quality factor comparable to plasmons in Au thin-films and hyperbolic PhPs in hBNs type-I Restrahlen band. It must be mentioned that hBN being a hyperbolic material, the physics which underlies its negatively dispersing polaritons is fundamentally different than our coupled mode model and thus may not be appropriately described by our current approach.

As discussed in the main text, the drawback of RB2 is the bandwidth of its asymmetric SPhP (defined as $\Delta\omega_- = \omega_{LO} - \omega_{sp,t}$). In fact, while measurements confirm a higher quality factor for the symmetric mSPhPs within the 2nd reststrahlen band, negative dispersion has yet to be realized, likely due to the narrow bandwidth of the asymmetric SPhP mode. Fortunately, interfacial coupling enables us to expand the range of negative dispersion simply by encapsulating the membrane in a high-dielectric material (fig. S6a). A more intuitive understanding of this bandwidth-tuning is gained upon inspecting the Rabi formula $\Delta = \sqrt{\left(\omega_{sp,t} - \omega_{sp,b}\right)^2 + 4\Delta_0^2 e^{-2qd}}$ where, for a symmetric dielectric environment, the two SPhP dispersions split about a common $\omega_{sp}$. Acknowledging that $\omega_{sp}$ takes on its maximum value at $\varepsilon_0 = 1$, the bandwidth $\Delta\omega_-$ must increase upon increasing $\varepsilon_0$. Fig. S6b plots the $\Delta\overline{\omega}_-$ as a function of $A$ and $\varepsilon_0$ (the surrounding dielectric environment). We also provide normalized bandwidths $\Delta\overline{\omega}_- = \frac{\Delta\omega_-}{g_{TO}}$ for several suspended materials in **Supplementary Table 2**. On a similar note, as a direct consequence of lowering $\omega_{sp}$, the group velocity $v_g$ and group index $Q_G$ are improved by a similar factor to $\Delta\overline{\omega}_-$. (Fig. S6c). Obviously, if we are to take $\Delta\overline{\omega}_-$ and $Q_G$ as indicators of the polariton "fitness", those materials boasting large oscillator strength (i.e. SiC, STO (RB2), hBN, and Au) hold the greatest potential.

# Supplementary References

## Supplementary Table 1: Measured samples organized by membrane thickness and substrate permittivity

| Substrate Permittivity / Membrane Thickness | Air ($\varepsilon = 1$) | $SiO_2$ ($\varepsilon = 3.9$) | Si ($\varepsilon = 11.7$) |
|---|---|---|---|
| 15 nm | | ✓ | |
| 25 nm | ✓ | ✓ | ✓ |
| 50 nm | ✓ | ✓ | ✓ |
| 120 nm | | ✓ | ✓ |

## Supplementary Table 2: Normalized lorentz oscillator parameters and figures of merit for thin-films (d = 100nm) of common polaritonic materials

| | STO (RB3) [30] | STO (RB2) [30] | SiC | AlN | TiO2 | Au | Ag | hBN |
|---|---|---|---|---|---|---|---|---|
| $A$ | 32.7 | 110.4 | 111.4 | 16.9 | 30 | 127.2 | 28.6 | 124.2 |
| $\bar{\omega}_0$ | 31.1 | 43.8 | 158.6 | 19.2 | 22.7 | 0 | 0 | 194.3 |
| $Q_\omega$ | 177 | 560 | 798 | 110 | 148 | 992 | 126 | 1083 |
| $Q_G$ | 1.3 | 3.5 | 2.1 | 1.1 | 1.3 | 4 | 1.5 | 2.9 |
| $\Delta\bar{\omega}_-$ | 2.0 | 8.6 | 5.2 | 1.0 | 2.0 | 10.7 | 2.4 | 5.5 |

# Supplementary Figures

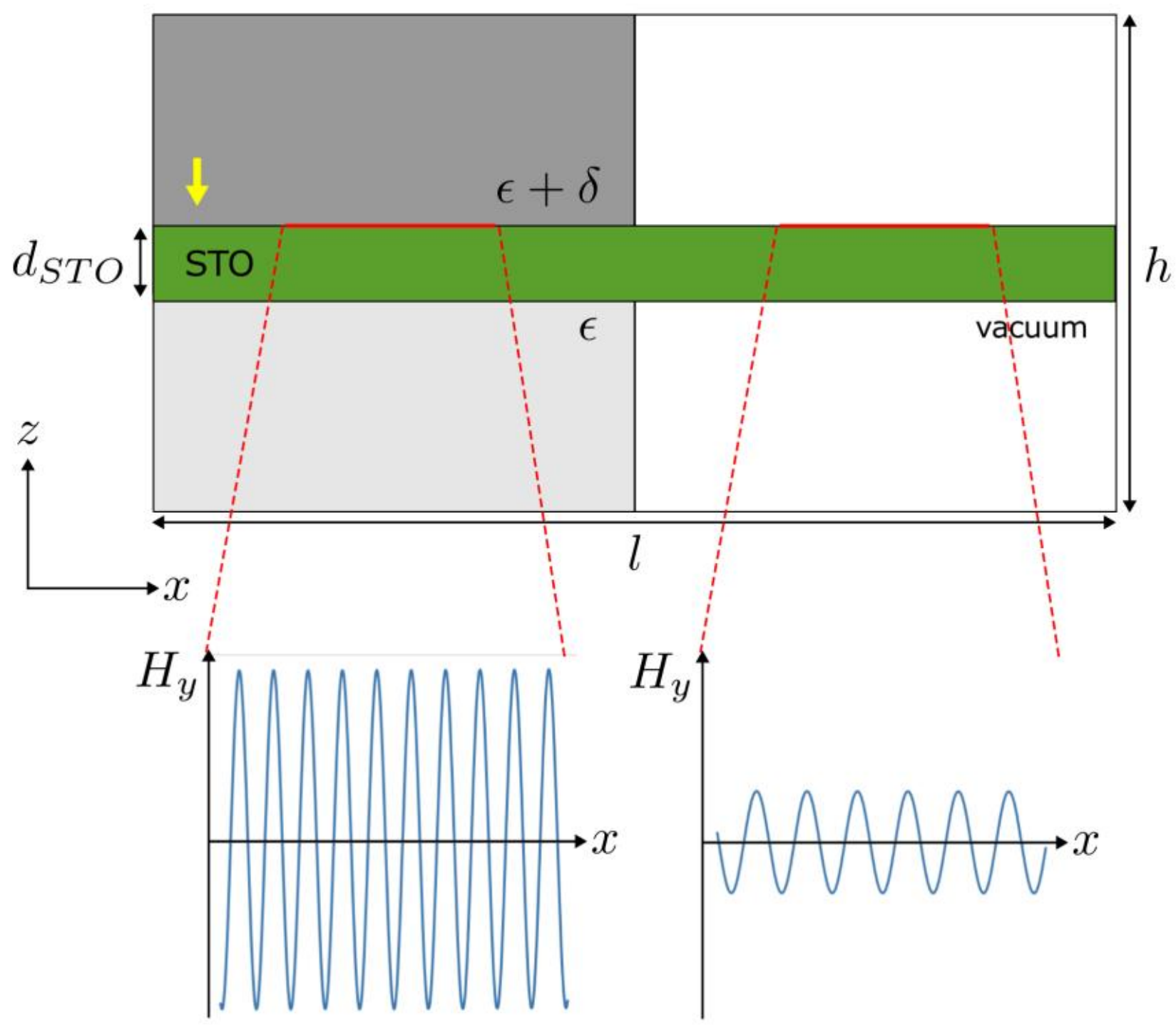

Figure S1 | Schematic representation of simulation domain for transmission across Veselago interface. Note that dimensions are not to scale. The $H_y$ field plots show the fields at the surface of the STO membrane in the two domains.

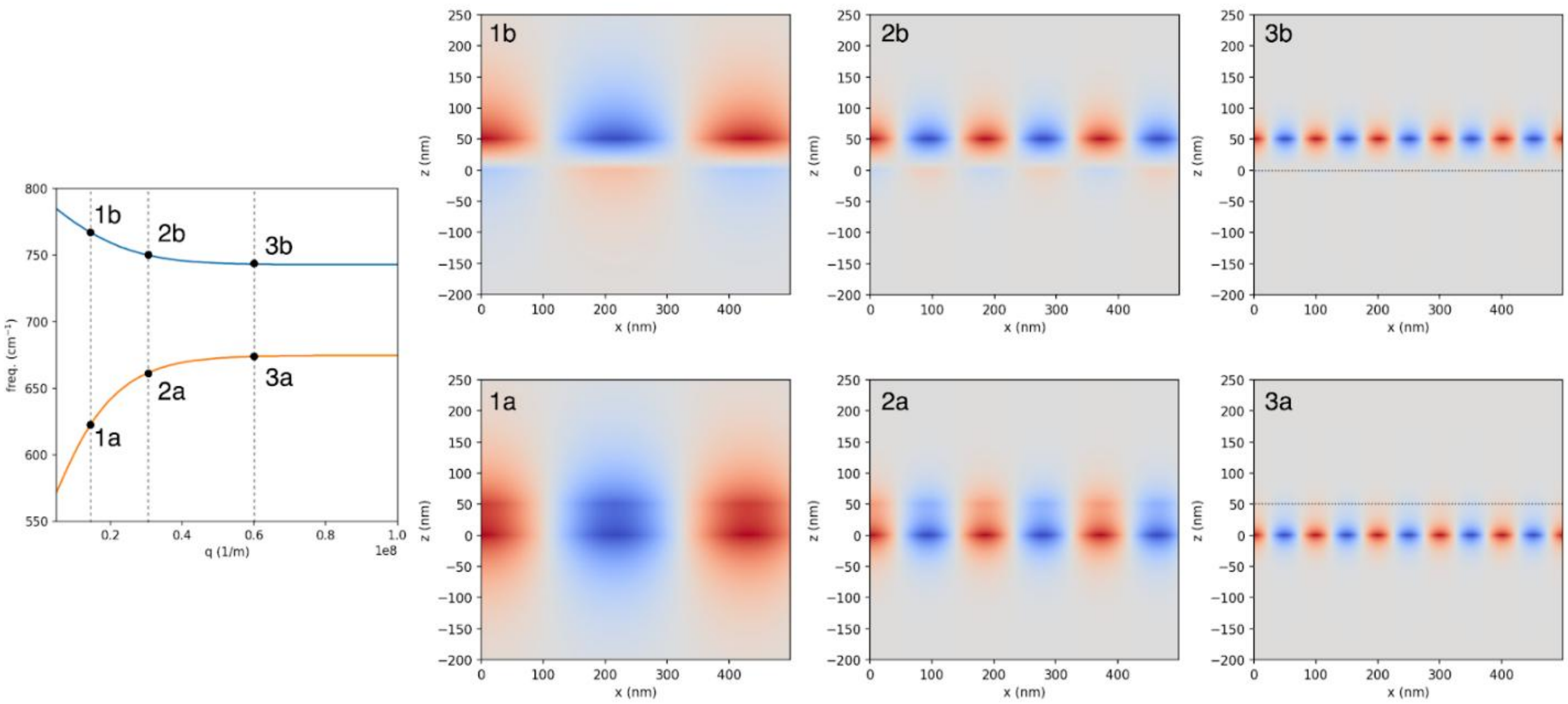

**Figure S2 | Dispersion and eigenmodes of STO slab in an asymmetric dielectric environment.** The substrate has dielectric constant $\varepsilon = 3$ while the superstrate has $\varepsilon = 1$. The STO slab thickness is d=50nm. (1b)-(3a) display the $E_x$ field associated quasi-symmetric (orange) and quasi-antisymmetric (blue) branches of the coupled polariton dispersion (left). The addition of an asymmetric dielectric environment to the membrane clearly breaks the otherwise perfectly symmetric mode profile that would be realized in Figs. 1b-2b in a symmetric dielectric environment. This feature is key to successful tunneling between modes, e.g. from (1a) to that in (1b), necessary for the Veselago lensing proposal presented in the main text.

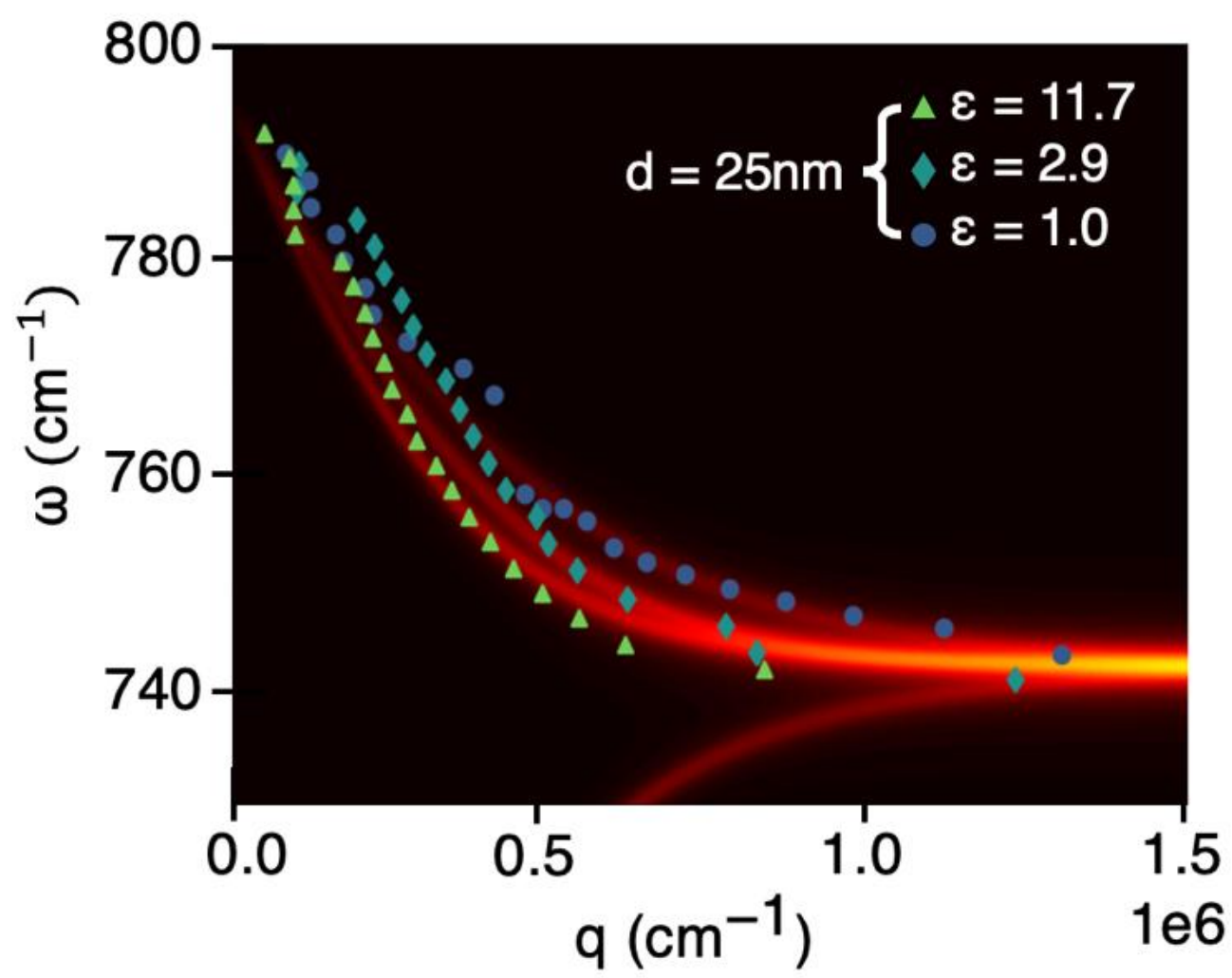

Figure S3 | Shift of asymmetric mode with substrate dielectric screening. Data points indicate measured momenta of the asymmetric polariton mode for a 25 nm STO membrane suspended, supported on silicon oxide, and supported on silicon, respectively, with corresponding to substrate permittivities $\varepsilon = 1$, 2.9, and 11.7 respectively. To the extent possible by measurement uncertainty, screening by higher permittivity substrates reduces the asymmetric polariton momentum $q_p(\omega)$ or, equivalently, elongates the asymmetric polariton wavelength (increases $\lambda_p(\omega)$).

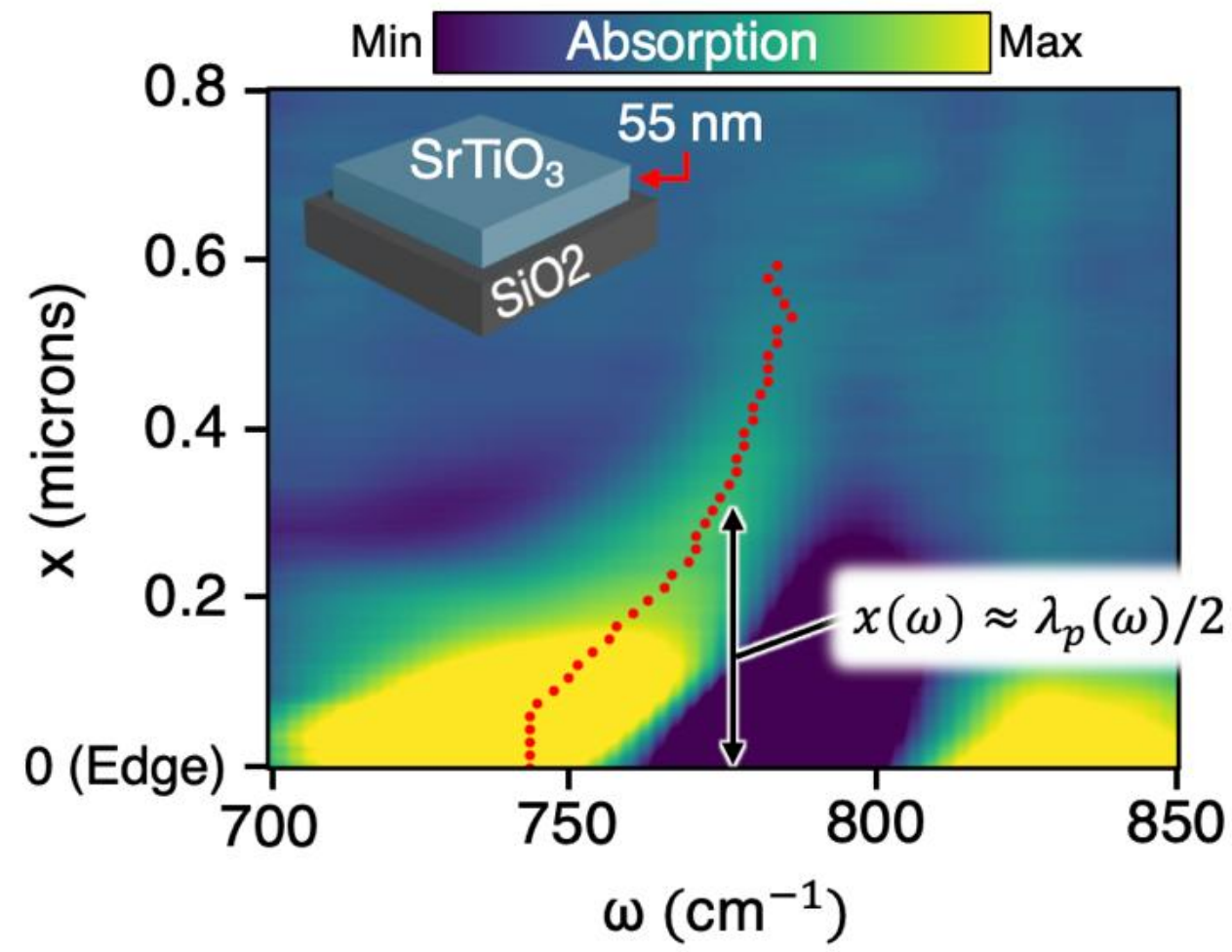

Figure S4 | Extraction of polariton dispersion from a nano-absorption linescan. A nano-resolved absorption linescan recorded on a 55nm thick STO membrane on SiO2 substrate spans its edge to 0.8 microns into its interior and resolves a dispersing position of maximum absorption which can be identified as the probe-edge distance producing constructive interference of a membrane phonon polariton. At each probe-edge distance $\boldsymbol{x}$, red data points identify the energy $\boldsymbol{\omega}$ at which absorption is maximized. As described in the main text with a simple model of polariton interferometry, these positions $\boldsymbol{x(\omega)}$ can be identified as $\boldsymbol{\lambda_p(\omega)/2}$, from which momenta $\boldsymbol{q_p(\omega) = 2\pi/\lambda_p(\omega)}$ are obtained and recorded as polariton dispersions presented throughout.

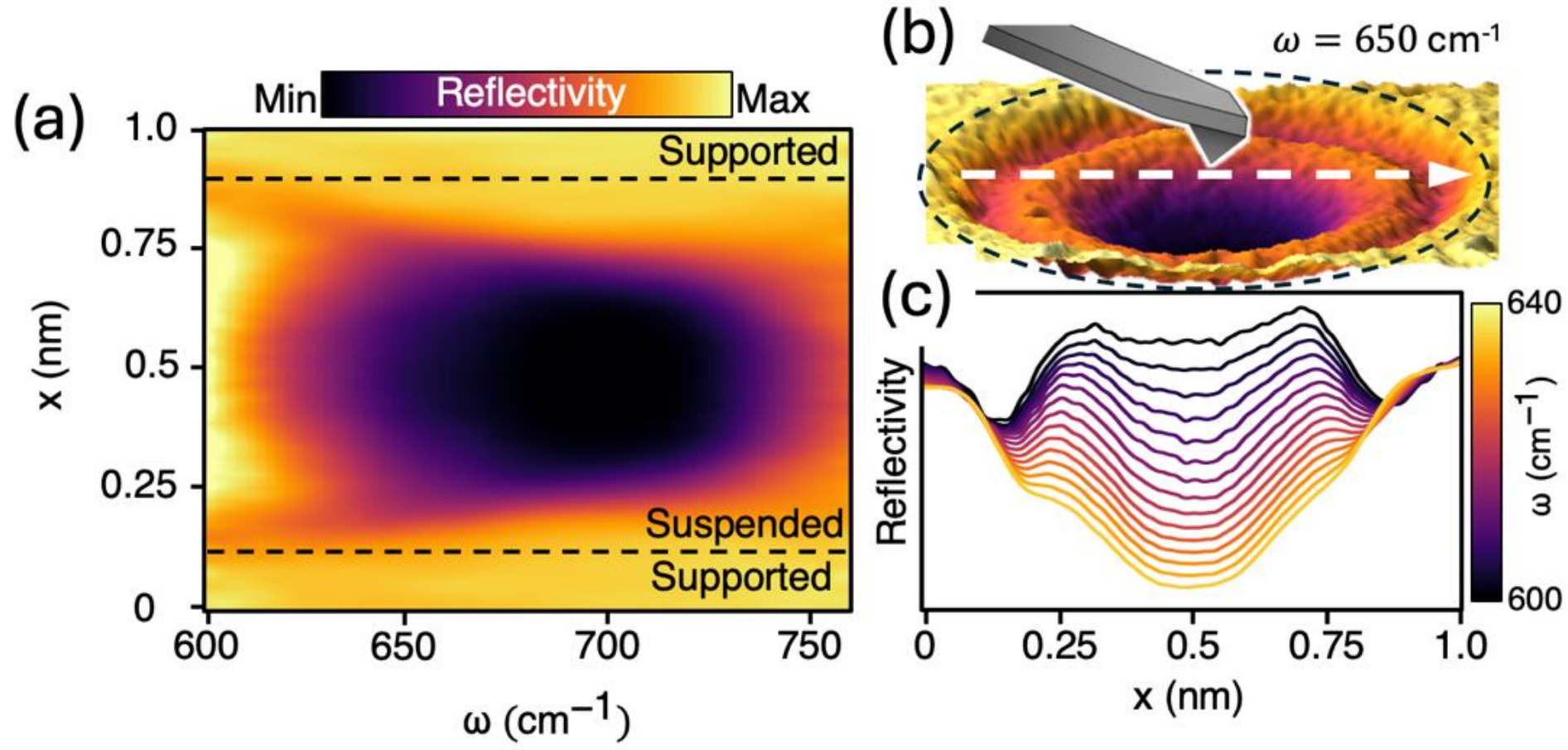

Figure S5 | Polariton interferometry across a suspended membrane "drumhead". (a) Nano-resolved linescan of reflectance from the same suspended membrane showcased in Fig. 4g (main text) resolved at energies from $\omega = 600$ cm-1 to 760 cm-1. (b) Height-rendered nano-scale reflection contrast at $\omega = 650$ cm-1 shows the inner "ring" arising from polariton interferometry internal to the suspended "drumhead" that is enhanced at its edge. (c) At increasing energies, the position of this fringe detectably moves "outward" toward the physical boundary of the suspended region, corresponding to a decreasing polariton wavelength $\lambda_p(\omega)$ associated with the (lower) symmetric membrane surface phonon polariton.

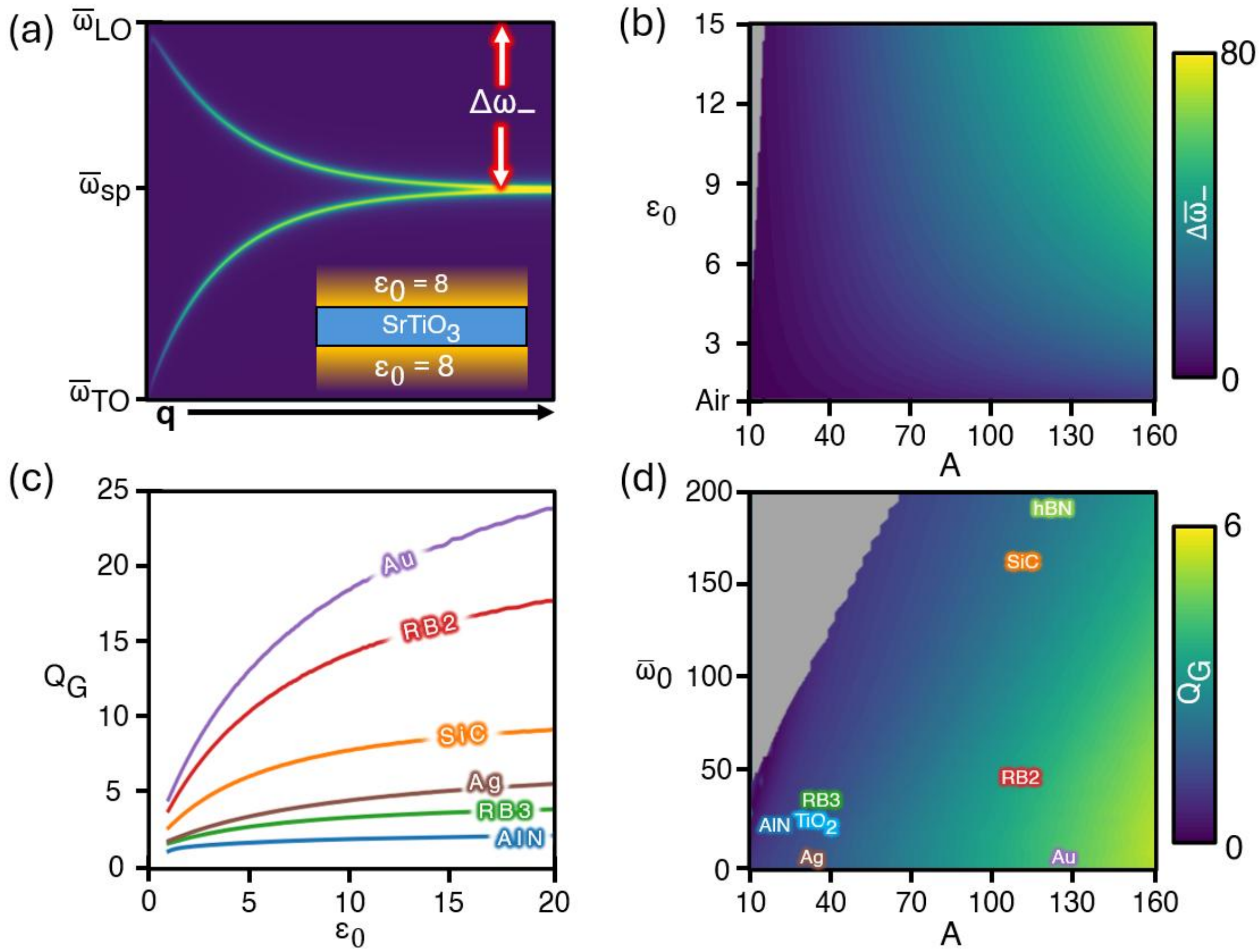

Figure S6 | Quality factor and bandwidth calculations for popular polaritonic materials. **(a)** Loss function calculations for the third Reststrahlen band (RB3) of a 100nm STO membrane encapsulated with a high-permittivity material. **(b)** Calculations of the bandwith of the $\omega_{-}$ polariton as a function of $A$ and $\varepsilon_0$. **(c-d)** High-permittivity environments allow for drastic improvements to the group quality factor ($Q_G$) in several popular polaritonic materials defined by $(A, \overline{\omega}_0)$.

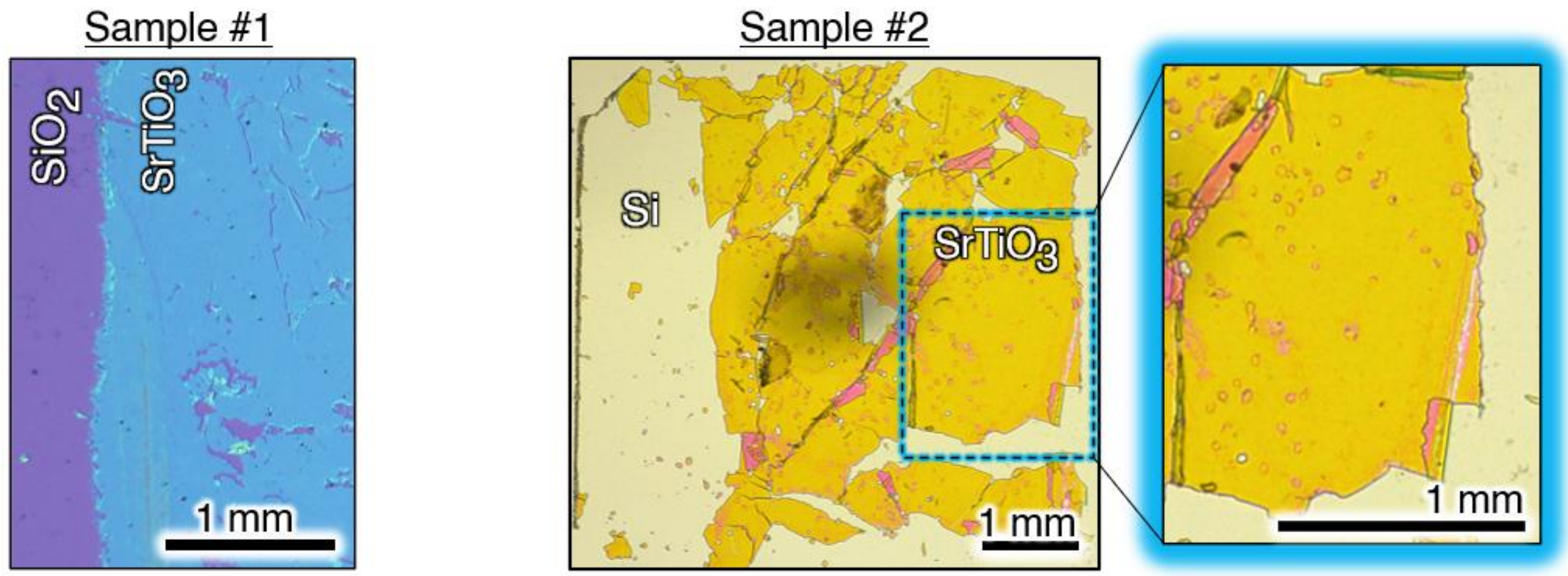

Figure S7 | Examples of large-area STO membranes on host substrates.

Large area membranes of thickness 5nm (left) and 120nm (right) after transfer to their respective host substrate.

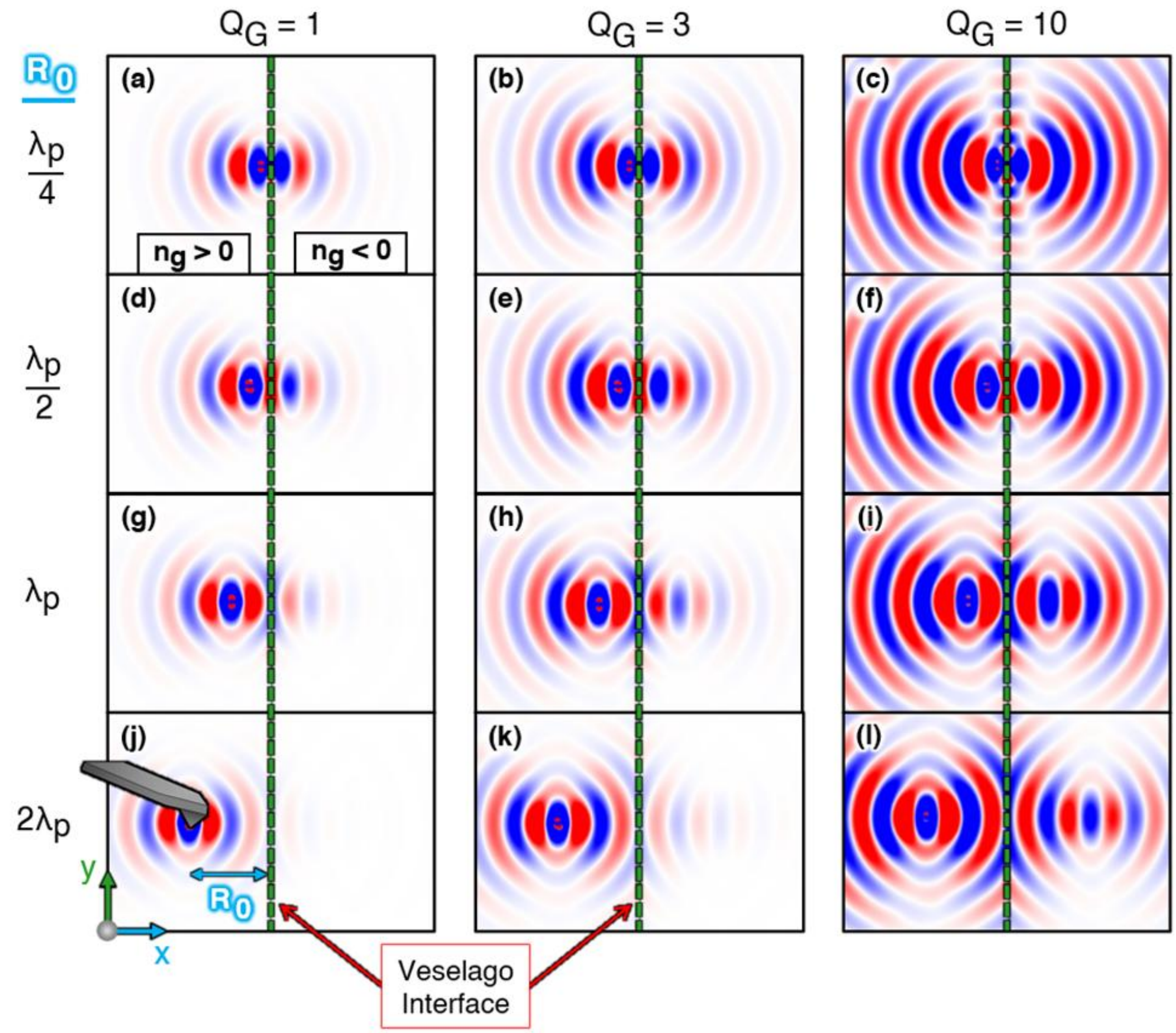

Figure S8 | Simulations of Veselago lensing for SPhPs with low quality factor. Real-space propagation patterns for polaritons with quality factors 1 (left), 3, (middle) and 10 (right). A Veselago focal point is clearly identified for all $Q_G$ at source-interface distances of ***a-c)*** $R_0 = \frac{\lambda_p}{4}$, ***d-f)*** $R_0 = \frac{\lambda_p}{2}$, and ***g-i)*** $R_0 = \lambda_p$. **j-l)** Further from the Veselago interface ($R_0 = 2\lambda_p$) focal points are identifiable for $Q_G \geq 3$.